\documentclass[prl,amsmath,twocolumn,longbibliography]{revtex4-1}

\usepackage[normalem]{ulem}
\usepackage{epsfig}
\usepackage{amsfonts}
\usepackage{amsmath}
\usepackage{color}
\usepackage{hyperref}
\setcitestyle{square}
\usepackage{subfigure}
\usepackage{natbib}
\usepackage{graphicx,amssymb}

\usepackage{comment,xcolor}

\begin{document}
\title{
Evolutionary Dynamics in a Varying Environment:  Continuous versus Discrete Noise
}
\author{Ami Taitelbaum\textsuperscript{a}}
\author{Robert West\textsuperscript{b}}
\author{Mauro Mobilia\textsuperscript{b,}}
\thanks{To whom correspondence should be addressed:  m.mobilia@leeds.ac.uk, michael.assaf@mail.huji.ac.il} \author{Michael Assaf\textsuperscript{a,c,}}
\thanks{To whom correspondence should be addressed:  m.mobilia@leeds.ac.uk, michael.assaf@mail.huji.ac.il}

\affiliation{$^a$Racah Institute of Physics, Hebrew University of Jerusalem, Jerusalem 91904, Israel}
\affiliation{$^b$Department of Applied Mathematics, School of Mathematics, University of Leeds, Leeds LS2 9JT, United Kingdom}
\affiliation{$^{c}$Institute of Physics and Astronomy, University of Potsdam, Potsdam 14476, Germany}

\begin{abstract}
Environmental variations can significantly influence how populations  compete for resources, and hence shape their evolution. Here, we study population dynamics subject to a fluctuating environment modeled by a varying carrying capacity changing continuously in time according to either binary random switches,  or by being driven by a noise of continuous range. We consider a prototypical example of two competing strains, one growing slightly slower than the other, and consider also  the scenario where the slow strain is a public goods producer. By
systematically comparing the effect of binary- versus continuously-varying environment, we study how different  noise statistics (mean,  variance) influence the population size and fixation properties. We show that the slow strain fixation probability can be greatly enhanced for a continuously-varying environment compared to binary switches, even when the first two moments of the carrying capacity
coincide.
\end{abstract}

\maketitle

	
Natural populations face endlessly varying environmental conditions,
such as the abundance of nutrients or toxins, temperature, light, and humidity, all of which influence their interactions and evolution \cite{Morley83,Fux05,Caporaso11}.
In the absence of detailed knowledge of how external factors change, they are often modeled as environmental noise (EN). This in turn shapes the fluctuating environment where populations evolve, for which several response mechanisms have been proposed~\cite{May73,Karlin74,Chesson81,Thattai04,Kussell05b,Assaf08,Assaf09,Loreau08,Beaumont09,Visco10,May73,Karlin74,He10,Tauber13,Assaf13,AMR13,Chisholm14,Kessler14,Kalyuzhny15,Assaf15,Melbinger15,Xue17,Assaf17,Assaf18,Dobramysl18,Marrec20}.
Apart from  EN, demographic noise (DN) is another  source of randomness: it can lead to fixation, when one species takes over the population,
and its effect is significant in small communities, but  negligible
in large populations~\cite{Kimura,Ewens,Blythe07,Nowak}.

Importantly, the evolution of the size and composition of a multispecies population are often interdependent~\cite{Roughgarden79,Leibler09,Melbinger2010,Cremer2011,Cremer2012,Melbinger2015a,Gokhale16,KEM1,KEM2}.
This may result in a coupling between DN and EN, with
external factors affecting the population size, which in turn modulates the DN intensity. The interplay between EN and DN
is crucial in microbial communities, which can experience sudden, extreme environmental changes~\cite{Wahl02,Rainey03,Patwas09,Wienand15,Brockhurst07a,Brockhurst07b,Coates18,Cremer19}, as well as in ecology~\cite{Chisholm14,Kessler14,Kalyuzhny15}. In the context of antimicrobial resistance,
variations of population size and composition are key
when antibiotics reduce a large community to a very small size, but fail to eradicate it. Surviving cells in the small population, prone to fluctuations,
may then replicate and restore infections, with survivors likely to develop antibiotic resistance
~\cite{Coates18,Marrec20}.  Interactions between microbial communities and the environment can also lead to population bottlenecks,
where new colonies of few individuals result in  cooperative behavior
~\cite{Rainey03,Brockhurst07a,Brockhurst07b}.
In most theoretical studies involving multiple species, there is no explicit interdependence between EN and DN. Growth rates are thus commonly assumed to be subject to noise of continuous range~\cite{May73,Karlin74,Kamenev08,AMR13,Assaf15,Melbinger15}, and vary independently of the population size that is often constant~\cite{Thattai04,Kussell05,Acar08,Gaal10,Assaf08,Assaf09,He10,Assaf13,Assaf15,Ashcroft14,Kussell05b,Melbinger15,Hufton16,Danino18,WMR18,Hufton18,Assaf18}.
On the other hand, there have been numerous lab-controlled experiments with microbial communities of varying sizes
evolving by switching instantaneously between a {\it discrete} number of environmental states
(``discrete EN''), with a strong focus on the binary case
\cite{Thattai04,Balaban04,Brockhurst07a, Acar08,Leibler09,Wienand15,Cremer19}. This has motivated the  study of
 population models with random binary  switching of the species
 growth rates~\cite{Thattai04,Visco10,Kalyuzhny15,Hufton16,Danino18,WMR18}, and more recently of the carrying capacity (or resources) leading
 to the coupling of DN and EN~\cite{KEM1,KEM2,WM19,TWAM,SMM}.

Nevertheless,
{\it in vivo} exogeneous factors often vary continuously, in time and over a range of values~\cite{Cavicchioli19,Nguyen21}, rather than by instantaneous switches. For instance,
the  carrying capacity of certain phytoplankton species
and the growth rates of some  algae
vary with the fluctuating temperature~\cite{Savage04,Bernhardt18,Descamps05}. It is thus important to understand how the
coupling of DN and EN affects the dynamics of communities in an
environment varying along a continuum
of states (``continuous EN''), and to compare its properties
with those  in binary fluctuating environments  commonly used in experiments~\cite{Thattai04,Balaban04,Brockhurst07a, Acar08,Leibler09,Wienand15,Cremer19} and theory~\cite{Visco10,Kalyuzhny15,Hufton16,KEM1,KEM2,WMR18,Danino18,WM19,SMM}.

Here, we address these questions by systematically investigating the influence of coupled DN and EN
on the evolution of a simplified microbial community consisting of two
competing strains, one growing slightly slower than the other,  subject to a carrying capacity driven  either by binary or  continuous
EN.
For this simplified microbial model, we
 unveil the similarities and differences of evolving
under continuous or binary EN, and reveal the drastic effect that  EN may have on
the population size distribution and fixation properties.
Remarkably, we show that the  slow species fixation probability
can be significantly enhanced under continuous noise  over its value under
binary EN of same mean and variance.


We consider a well-mixed population consisting of $N_S$  individuals of a
slow-growing strain $S$  and $N_F$ microbes of the fast-growing strain $F$.
At time $t$, this two-strain population has a time-fluctuating size $N(t)=N_S(t)+N_F(t)$
and is composed of a fraction $x=N_S/N$ of slow growers $S$.
 Per-capita growth rates are $(1-s)/\bar
f$ for
$S$   and $1/\bar f$ for $F$,
where
$\bar f=1-sx$ is the population average fitness,
and $0<s\ll1$ denotes the small growth advantage (selective bias) of  $F$ over $S$~\cite{Melbinger2010,Cremer2011,KEM1,KEM2,TWAM}.  Owing to limited and varying resources,
the strain's growth is limited 	by a logistic death rate $N/K_\ell(t)$, where $K_\ell(t)\gg 1$ is the carrying capacity that here fluctuates in time due to EN.
 This  allows us to couple in a simple and biologically-relevant way
 DN and EN~\cite{KEM1,KEM2,WM19,TWAM,SMM},
which yields the following birth-death process~\cite{Gardiner,KEM2}:
		\vspace{-2.8mm}
\begin{eqnarray}
	\label{eq:BDP}
	N_{S/F}  \xrightarrow{T_{S/F}^{+}}  N_{S/F}+ 1 \; \text{and} \;
	N_{S/F}  \xrightarrow{T_{S/F}^{-}}  N_{S/F}- 1,
\end{eqnarray}
with transition rates $T_{S}^{+}= (1-s)N_S/\bar{f}$,  $T_{F}^{+}= N_F/\bar{f}$ and $T_{S/F}^{-}= (N/K_\ell)N_{S/F}$.
We model EN by letting the carrying capacity fluctuate in time as:
\begin{equation}
	\label{eq:K}
	K_\ell(t) =K_{0,\ell}[1+\xi_{\ell}(t)],
\end{equation}
where $\xi_{\ell}(t)$
 denotes the stationary symmetric EN of type $\ell\in{\cal L}\equiv \{{\rm D}, {\rm U},  {\rm B} \}$, with  discrete or continuous range.
 For the former, we focus on the {\it symmetric dichotomous} ($\ell={\rm D}$, or {\it telegraph}) noise~\cite{Bena06,HL06}, and for the latter we consider EN with uniform
  ($\ell={\rm U}$) and symmetric beta ($\ell={\rm B}$) stationary probability density function (PDF) $p^*(\xi_{\ell})$ of support ${\cal S}_{\ell}$,
  see Supplemental Material (SM)~\cite{SM}. It is convenient to denote the set of EN with continuous range (continuous EN) as $\ell \in {\cal L}'\equiv {\cal L}\setminus {\rm D}$.

\begin{figure}[t]
	\centering
	\includegraphics[width=1\linewidth]{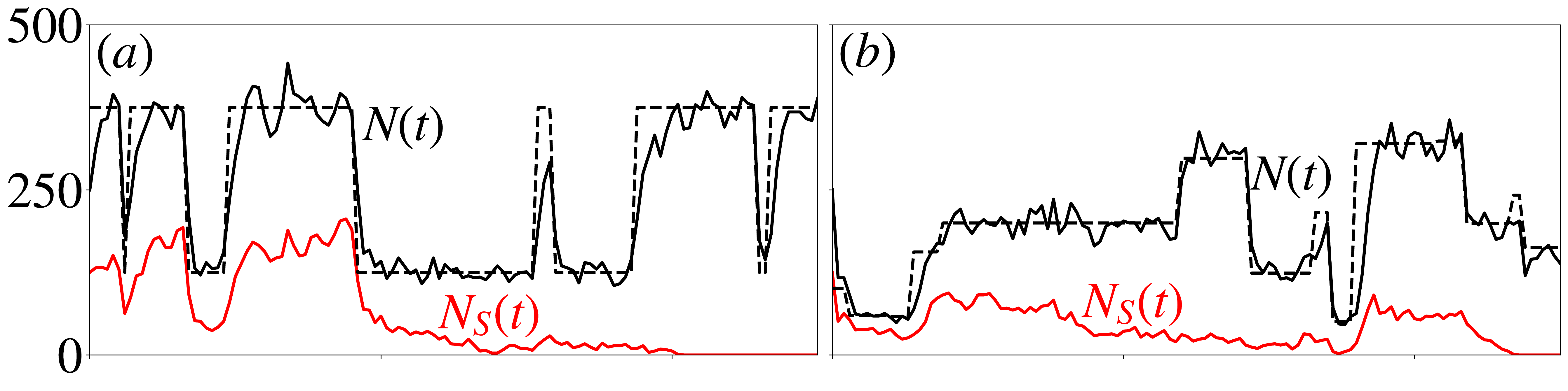}
	\vspace{-1.2mm}
    \includegraphics[width=1\linewidth]{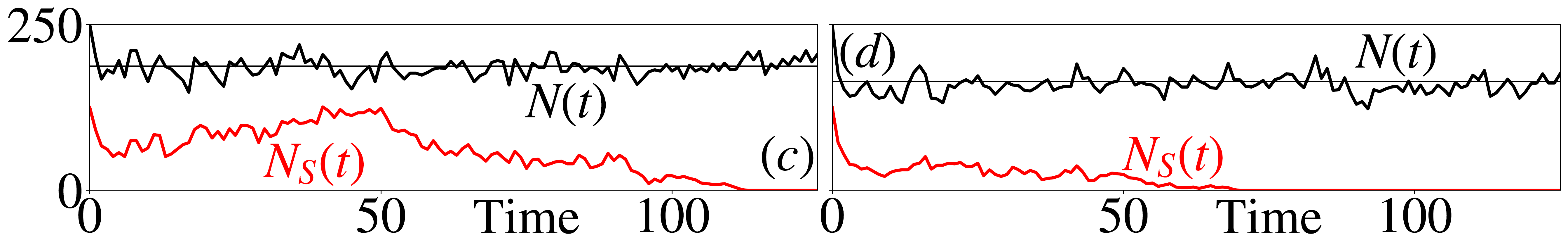}
	\vspace{-3mm}\caption{
	Typical realizations of $N$ (black) and $N_S$ (red/gray) vs. time under symmetric EN of variance $\sigma^2$: binary ${\rm D}$-noise in (a,c), and continuous ${\rm U}$-noise in (b,d).
	(a,b)
	$\nu = 0.1$, dashed lines show $K(t)$. (c,d) $\nu=1000$, solid lines show $\mathcal{K}_{\rm D}$ (c) and $\mathcal{K}_{\rm U}$ (d)  given by Eq. \eqref{eq:curlyK}, see text.
	In all panels $(s, x_0,K_0, \sigma) = (0.02, 0.5, 250, 0.5)$, and  $N(0)=K_0$.
	}\label{fig:popaths}
\end{figure}

For $\ell={\rm D}$, the random process
 $\xi_{\rm D} \!\to\! -\xi_{\rm D}$, where $\xi_{\rm D} \in \lbrace -\sigma_{\rm D} , \sigma_{\rm D} \rbrace$ ($0\!<\!\sigma_{\rm D}\!<\!1$), occurs  at rate  $\nu/2$ and has   correlation time
$1/\nu$. Hence, $K_\ell$  switches
between a high and low value  after an average time
$2/\nu$, see Fig. \ref{fig:popaths}.

For $\ell \in {\cal L}'$, $\xi_{\ell}$ is a colored continuous EN with correlation time $1/\nu$
  defined by the stochastic differential equation, in the sense of It\^o calculus~\cite{Gardiner,SM}:
	\begin{equation}
	\label{eq:SDE}
			d\xi_{\ell} = -\nu\xi_{\ell}dt  + \sqrt{{\cal B}_{\ell} } ~dW,
		\end{equation}
	where  $W\sim {\cal N}(0,1)$ is the normally-distributed Wiener process of zero mean and unit variance. The first term  on the right-hand-side of  \eqref{eq:SDE} represents the linear drift, and
the second  is the diffusion term. For concreteness and simplicity, we focus on symmetric continuous EN: $\ell={\rm B}$ (beta)
and $\ell={\rm U}$ (uniform) as examples of EN for which $K_{\ell}$ has a zero and finite lower bound, respectively. In the former,
$\xi_{\rm B}$ is distributed according
to a single-parameter  ($\beta>1$) symmetric beta distribution on $(-1,1)$, with variance
$\sigma_{\rm B}^2=1/(2\beta +1)<1/3$; in the latter
$\xi_{\rm U}$ is uniformly distributed   on  ($-\sigma_{\rm U}\sqrt{3},\sigma_{\rm U}\sqrt{3}$) with variance $\sigma_{\rm U}^2<1/3$, see Fig. \ref{fig:popaths}.
The  diffusive terms satisfy~\cite{SM}
\begin{equation}
{\cal B}_{{\rm U}}=
\nu(3\sigma_{{\rm U}}^2-\xi_{{\rm U}}^2),\;\;\;\;\;\;
{\cal B}_{{\rm B}}=\nu\left(\frac{2\sigma_{{\rm B}}^2}{1-\sigma_{{\rm B}}^2}\right)(1-\xi_{{\rm B}}^2).
\end{equation}
Notably, the coupling of \eqref{eq:BDP}-\eqref{eq:SDE}
 generally yields a {\it non-Markovian} process (when $\nu\neq 0$), see Sec.~A2 of \cite{SM}.

The PDF of $K_{\ell}$, ${\cal P}(K_{\ell})$, can  be obtained from $p^*(\xi_{\ell})$ and Eq.~\eqref{eq:K}~\cite{SM}. Below we focus on
the first two moments of $K_{\ell}$ (skewness vanishes for symmetric EN). To meaningfully compare the influence of
discrete and continuous EN on population dynamics,
we impose the same first two moments, yielding $K_{0,{\rm D}}\!=\!K_{0,\ell}\!=\!K_0$ and $\sigma^2\!=\!\sigma_{\rm D}^2\!=\!\sigma_{\ell}^2$ for $\ell\in {\cal L}'$, as long as $\sigma_{\ell}^2<\sigma_{\rm max}^2\equiv1/3$~\cite{SM}. Henceforth, as long as $\sigma_{\ell}<\sigma_{\rm max}$ we denote $\sigma_{\ell}$ by $\sigma$ for all forms of EN.

Ignoring fluctuations, in the limit of an infinite population with constant carrying capacity $K\left(t\right)=K_0\gg 1$,  the resulting mean-field dynamics yields $\dot{N}=N(1-N/K_0)$ and $\dot{x}\approx -sx(1-x)$~\cite{Melbinger2010,KEM1,KEM2,TWAM}. This indicates a timescale separation between the typical relaxation time of $N$, $t={\cal O}(1)$, and that of $x$, $t\sim 1/s\gg 1$.
Accounting for DN, the above timescales represent the convergence of the population size distribution (PSD) to the long lived metastable state  centred about $K_0$, after  $t={\cal O}(1)$, and the fixation of one of the species (and extinction of the other), at $t\sim 1/s$~\cite{KEM1,KEM2,SM}~\cite{time_scales}.

Indeed, in a finite population, random birth/death events  lead to the  fixation of one  strain.
The slow-grower
 fixation probability in a population of constant size $N$, given  an initial $x_0=N_S(0)/N(0)$, satisfies:
 $\phi(N,s,x_0) = (e^{-Nx_0\ln(1-s)}-1)/(e^{-N\ln(1-s)}-1)\approx
e^{N(1-x_0)\ln{(1-s)}}$, where the  approximation holds when $-N\ln(1-s)\simeq Ns\gg 1$~\cite{Ewens,Antal}, and when $N$ fluctuates about $K_0\gg 1$~\cite{KEM1,KEM2,TWAM}. However, the fixation probability changes dramatically when
$K_\ell$ varies according to \eqref{eq:K}.
 Since EN
 varies either discretely or continuously,
 we characterize the population dynamics by studying the joint influence of EN and DN  on the fixation properties
 and PSD as function of $\nu$ and $\sigma$.

In the case of {\rm D}-EN, the full PSD, $P_{\ell}(N,\nu)$, can be
well approximated
in all regimes by the PDF of a piecewise-deterministic Markov process associated with \eqref{eq:BDP}-\eqref{eq:K}~\cite{Davis84,KEM1,KEM2,WM19,TWAM,SMM,SM}.
Yet, there is no equivalent method to approximate the PSD for all $\nu$ under continuous EN, and  $P_{\ell}(N,\nu)$ is thus obtained numerically, see Sec.~A3 and Fig.~S1 in~\cite{SM}. As detailed below, analytical progress is however possible in the regime $\nu \ll s$
 (long correlation time), when $K_{\ell}(t)\approx K_{\ell}(0)$ and $P_{\ell}(N,\nu/s)\approx {\cal P}(K_{\ell})$, and when $\nu \gg s$ (short correlation time). In the latter regime, $P_{\ell}(N,\nu/s)\approx \delta\left(N-\mathcal{K}_\ell\right)$, see Eqs.~\eqref{eq:curlyKfull},\eqref{eq:curlyK} and Fig.~S1(c) in~\cite{SM}. Furthermore, when $\nu \gg s$ and $\sigma\ll 1$, $P_{\ell}(N,\nu)$ can be computed more accurately within a WKB approximation, see Sec. A2.3 in \cite{SM}.
 Once the PSD  found, numerically or analytically, we can use the timescale separation to find the $S-$fixation probability under $\ell$-EN, $\phi_{\ell}$. Indeed, the system settles in its long-lived PSD after a time of ${\cal O}(1)$, while fixation occurs after a time ${\cal O}(1/s)\gg 1$.
Hence, given $x_0$,
$\phi_{\ell}$ can be found by averaging $\phi(N,s,x_0)$
 over  $P_{\ell}(N, \nu/s)$:
\begin{eqnarray}
 \label{eq:formula}
 \phi_{\ell}\simeq
 \int_{0}^{\infty}~P_{\ell}(N,\nu/s)~\phi(N,s,x_0)~dN,
\end{eqnarray}
 where we have rescaled $\nu \to \nu/s$~\cite{KEM1,KEM2,TWAM}. This result holds under weak selection,  $1/K_0\ll s\ll 1$,
when EN varies on average ${\cal O}(\nu/s)$
 times prior to fixation~\cite{KEM1,KEM2}.
 A similar approach allows us to obtain the mean fixation time $T_{\ell}={\cal O}(1/s)$, see Sec. A4 in ~\cite{SM}.
Before considering the general case using \eqref{eq:formula}, we now study the PSD and $\phi_{\ell}$ in the
regimes of long and short-correlated EN.

\par
\begin{figure}[h!]
	\centering
	\includegraphics[width=1\linewidth]{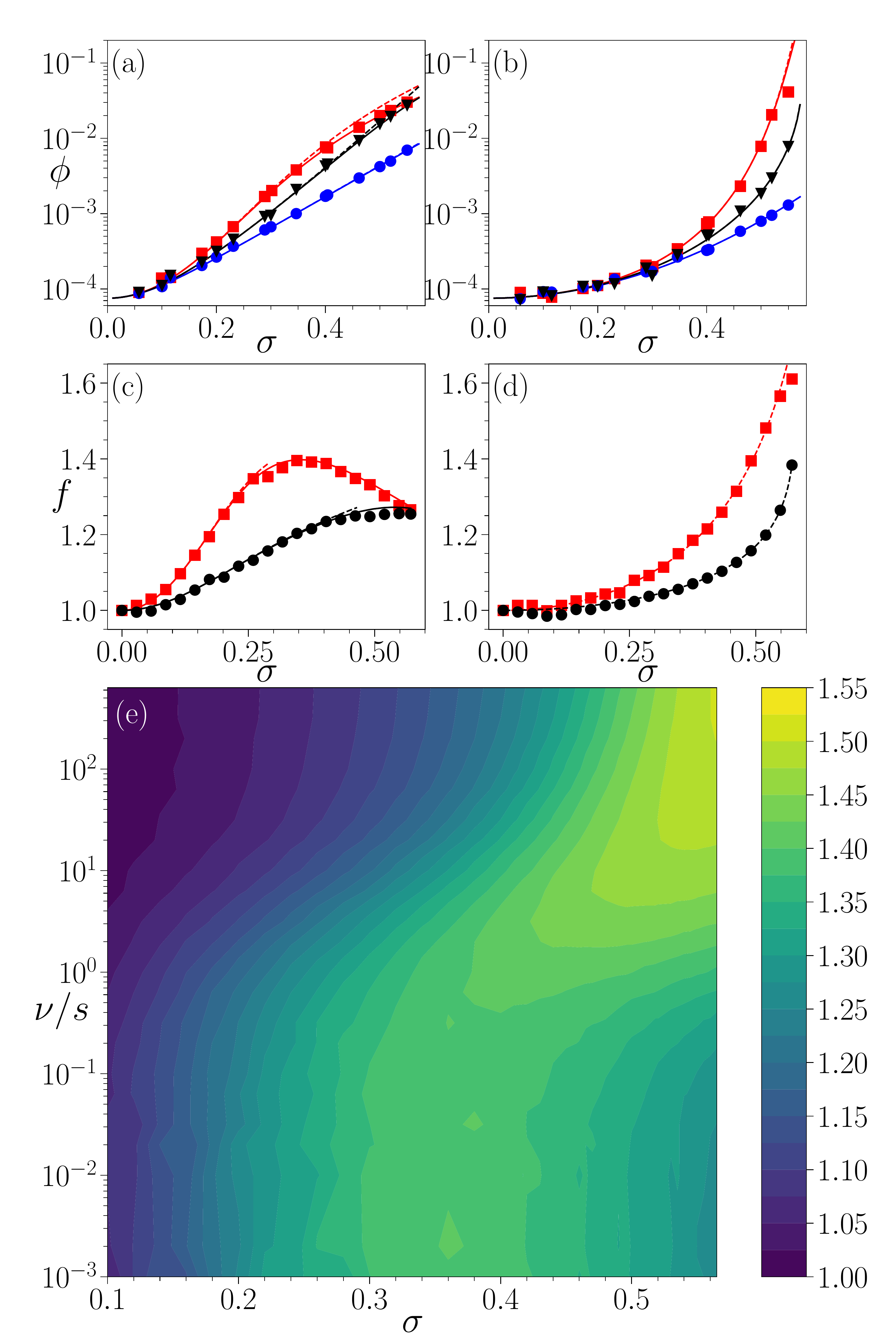}
	 \vspace{-5mm}\caption{(a) $\phi_{\ell}^{0}$ vs. $\sigma$ and (b) $\phi_{\ell}^{\infty}$ vs. $\sigma$
	under  symmetric $\ell$-EN.
	(c) $f_{\ell}^0$ vs. $\sigma$ and (d) $f_\ell^\infty$ vs.  $\sigma$, where $f_{\ell}$ is the standard-deviation multiplier
	obtained by solving
	$\phi_{\rm D}^{(0,\infty)}(f_{\ell}^{(0,\infty)}\sigma) = \phi_{\ell\in {\cal L}'}^{(0,\infty)}(\sigma)$
	with $K_0$ kept fixed, see text.
	 Symbols in (a,b) are from simulations, and in (c, d) are from numerical evaluation of \eqref{eq:formula} using simulation histograms for $P_{\ell}$. Dashed lines are from Eqs. \eqref{eq:phi-zero} in (a,c) and \eqref{eq:curlyK} in (b,d), while solid lines in (a)-(c) are from numerical evaluation of \eqref{eq:formula} using analytical results for $P_{\ell}$ under low/high $\nu$ limits. In (a)-(d): blue/dark gray for $\ell={\rm D}$, black for $\ell={\rm U}$
	and red/gray for $\ell={\rm B}$.
	 Simulation results for $\phi_{\ell}^{0}$ and  $\phi_{\ell}^{\infty}$
	 were obtained for  $\nu=10^{-4}$ and $\nu=3\cdot10^3$, respectively.
	(e) Heatmap of the multiplier $f_{\rm B}$ vs. $\sigma$ and $\nu$, see text. Dark areas interspersed by  lighter regions indicate where
	$f_{\rm B}(\sigma)$ is nonmonotonic (slow/intermediate $\nu$-regimes).
	In all panels $\left(K_0,s,x_0\right)=\left(750,0.025,0.5\right)$,
	and $\sigma <\sigma_{\rm max}$ for proper comparison of the different
    EN. A  similar heatmap is obtained for {\rm U}-EN, with the main qualitative difference being the absence non-monotonicity.
	}
	\label{fig:figure2}
\end{figure}

{\it Low varying rate (long-correlated EN).}
When $\nu\ll s$, the environment barely changes prior to fixation
of either species (after $t\sim 1/s$), and is assumed to be stationary as $N$ rapidly  equilibrates, with  $P_{\ell}\left(N\right) \approx {\cal P}\left(K_{\ell}\right)$, see Sec.~A2 in  \cite{SM}.  We thus approximate  $\phi_{\ell}$  by
$\phi_{\ell}^0 = \int
{\cal P}(K_{\ell})\phi(K_{\ell},s,x_0) dK_{\ell}$~\cite{SM}. Here,
the PSD is unimodal (or flat) under continuous EN,
in sharp contrast to the bimodal PSD obtained for {\rm D}-EN~\cite{KEM1}, see Fig.~S1(a,b) in  \cite{SM}.
When $K_0 s\gg 1$ and $s\ll 1$, we have
$\phi(N,s, x_0)\approx {\rm exp}(-\eta N)$  with $\eta\equiv -\left(1-x_0\right)\ln\left(1-s\right)\simeq s(1-x_0)>0$. By integrating over ${\cal P}(K_{\ell})$, we find
  \begin{eqnarray}\label{eq:phi-zero}
  \hspace{-4mm}
     \phi_{\ell}^0=
\begin{cases}
e^{-\eta K_0}\cosh{\left(\eta K_0\sigma\right)} & ({\rm D})\\

\int_{-1}^{1}  e^{-\eta K_0\left(1+\xi\right)} \frac{\left(1+\xi\right)^{\beta-1}\left(1-\xi\right)^{\beta-1}}{B\left(\beta,\beta\right)2^{2\beta-1}}~d\xi&  ({\rm B}),\\

\frac{e^{-\eta K_0}}{\eta K_0\sigma\sqrt{3}} \sinh \left(\eta K_0 \sigma \sqrt{3} \right)
 & ({\rm U})
\end{cases}
  \end{eqnarray}
where   $\beta \equiv \left(1-\sigma^2\right)/\left(2\sigma^2\right)$ and $B\left(\beta,\beta\right)\equiv\int_{0}^{1}t^{\beta-1}(1-t)^{(\beta-1)}~dt $ is the beta function.
Since the first two moments of the EN and $K_{\ell}$ coincide, $\phi_{\ell}^0$ depends only on $K_0$ and $\sigma$. In Fig.~\ref{fig:figure2}~(a) we show the dependence of $\phi_{\ell}^0$ on $\sigma$ for fixed $K_0$, which agrees well with simulation results. We find
that EN can enhance the $S$-fixation probability by several order magnitudes
with respect to  $\phi(K_0,s,x_0)$, its static-environment counterpart~\cite{KEM1,KEM2,SM}, see Fig. S3 of~\cite{SM}. Moreover, $\phi_{\ell\in {\cal L}'}^0$ under continuous EN is much larger than $\phi_{{\rm D}}^0$. This stems from $P_{\ell\in {\cal L}'}$ having a broad left tail enhancing $\phi_{\ell}^0$  over
 the contribution arising from the left peak of $P_{{\rm D}}$, see Eq.~\eqref{eq:formula} and Fig. S1(a,b) in~\cite{SM}.

{\it High varying rate (short-correlated EN).}
When $\nu \gg s$  and $K_0\gg 1$ (with $\sigma$ not too close to $\sigma_{\rm  max}$), $\phi_{\ell}$ is governed by EN that dominates over DN~\cite{TWAM,SM}.
In fact, under high $\nu$, $N$ obeys the  logistic stochastic differential equation $\, \dot{N}=N\left(
 1-N/K\left(t\right)\right)$, with the  environment varying so frequently that
 EN self-averages, yielding $\dot{N}=N\left(1-N/\mathcal{K}_{\ell} \right)$~\cite{KEM1,KEM2,TWAM},  where~\cite{SM}
\begin{eqnarray}
 \label{eq:curlyKfull}
 \mathcal{K}_{\ell}
 \equiv \frac{K_0}{\langle \frac{1}{1+\xi_{\ell}}\rangle}
 \equiv K_0 \left[\int_{{\cal S}_{\ell}} \frac{p^*(\xi_{\ell})}{1+\xi_{\ell}}~d\xi_{\ell}\right]^{-1}.
\end{eqnarray}
 For symmetric $\ell$-EN, we explicitly find
 	\begin{equation}
	\label{eq:curlyK}
	\hspace{-2mm}\frac{{\mathcal{K}_{D}}}{K_0}\!=\!             1\!-\!\sigma^2,\;\;\;
	  \frac{{\mathcal{K}}_{B}}{K_0}\!=\!1\!-\!	\frac{\sigma^2}{1\!-\!2\sigma^2},\;\;\;
	 \frac{{\mathcal{K}}_{U}}{K_0}\!=\! \frac{\sqrt{3}\sigma}{\tanh^{\!-1\!}\!\left(\sqrt{3}\sigma\!\right)}\!.
	\end{equation}
	This dependence
	yields ${\cal K}_{\rm B}<{\cal K}_{\rm U}<{\cal K}_{\rm D}$
	for fixed $\sigma$,  see Fig. S2(a) of~\cite{SM}.
	 $P_{\ell}\left(N,\nu/s\right)$ is  very narrow and centered around $\mathcal{K}_\ell$  when $\nu \gg s$, see Fig.~S1(c) of~\cite{SM}.
	Hence, upon ignoring DN, Eq. \eqref{eq:formula} can be crudely approximated using $P_{\ell}\left(N,\nu/s\right)\approx\delta\left(N-\mathcal{K}_\ell\right)$, yielding
  $\phi_{\ell}\to \phi_{\ell}^\infty\approx\exp\left(-\eta\mathcal{K}_{\ell}\right)$: when $\nu/s\gg 1$,
  high environmental variability ensures self-averaging prior to fixation, leading to $\phi_{\ell}\approx \phi_{\ell}^\infty$.
  According to  \eqref{eq:curlyK} the values of $\mathcal{K}_{\ell}$ for
 $\ell \in {\cal L}'$ are markedly lower than $\mathcal{K}_{\rm D}$, especially when $\sigma$ approaches $\sigma_{\rm max}$. This
 implies $\phi_{\ell \in {\cal L}'}^\infty \gg \phi_{\rm D}^\infty$, as confirmed by Fig.~\ref{fig:figure2}~(b), whose predictions  agree well with simulation data. Also,  $\phi_{\ell}^\infty$  is generally
 significantly larger than its static-environment counterpart, see Fig. S3~\cite{SM}.

 {\it Intermediate varying rate  (general case).}
  When $\nu  \sim s$, we
  compute $\phi_{\ell}$ using  \eqref{eq:formula}.
  While an analytical approximation can be obtained under {\rm D}-EN~
  \cite{KEM1,KEM2,WM19,TWAM},
  \eqref{eq:formula} is evaluated numerically
  under continuous EN by integrating over the PSD obtained from simulation data.
 We find that  \eqref{eq:formula} efficiently provides an accurate
 approximation of $\phi_{\ell}$ over a broad range of $\nu/s$ for all forms of $\ell$-EN, see Fig.~S3 of~\cite{SM}. This approximation agrees well with
 $\phi_{\ell}^{0}$ when $\nu/s\ll 1$ and $\phi_{\ell}^{\infty}$ when $\nu/s\gg 1$~\cite{SM}.

Most lab-controlled experiments on fluctuating populations are carried out
 by probing a discrete set  (often binary) of environmental states~\cite{Acar08,Thattai04,Balaban04,Kussell05,Abdul-Rahman21}. Yet, many {\it in vivo} exgoneous factors can take a  continuous range of values, and
 little is known on the joint effects of continuously-varying environmental conditions and DN. We thus  analyze  the effects of
 discrete and continuous EN on population dynamics, by systematically comparing $\phi_{{\rm D}}$  under {\rm D}-noise with $\phi_{{\rm B/U}}$ under {\rm B/U}-EN. Keeping  $K_0$ and $\sigma$  fixed
 for every $\ell$-EN, we have determined the multiplier $f_{\ell}$
  by which the variance of the {\rm D}-noise needs to be enhanced ($\sigma\to f_{\ell}\sigma$) for $\phi_{{\rm D}}$ to match $\phi_{\ell\in {\cal L}'}$,
 for  given $\nu$.
In practice, we have generally used \eqref{eq:formula}
  to determine $f_{\ell}$
 by numerically solving
 $\phi_{\rm D}(f_{\ell}\sigma) = \phi_{\ell}(\sigma)$ over $\sigma\in (0,\sigma_{\rm max})$
 for $\ell\in {\cal L}'$ and fixed $\nu$, see Sec.~A3 in \cite{SM}. As shown in Fig.~\ref{fig:figure2}~(c-e),
 $f_{\ell}$ is a nontrivial function of $\nu$ and $\sigma$, with $f_{\ell}\geq 1$ reflecting the fact that a higher variance of  {\rm D}-EN is necessary to achieve the same fixation probability as under {\rm B/U}-EN.
 For long- and short-correlated EN  ($\nu/s\ll 1$ and $\nu/s\gg 1$, respectively) we have used \eqref{eq:phi-zero} and \eqref{eq:curlyK} to determine  the multipliers $f_{\ell}^0$ and $f_{\ell}^\infty$ analytically. These  predictions, shown in  Fig.~\ref{fig:figure2}~(c,d),  agree well with simulation results. For $\sigma\ll 1$, in the limit $\nu\to 0$, one has
$ f_{\rm U}^0 \!\approx\!
	  1\!+\!(\eta K_0 \sigma)^2/30$, $f_{\rm B}^0 \!\approx\!
	  1\!+\!(\eta K_0 \sigma)^2/12$,  while
 $f_{\rm U}^\infty \!\approx\!
	  1\!+\!(2/5)\sigma^2$ and $f_{\rm B}^\infty \!\approx \!
	  1\!+\!\sigma^2$ when $\nu\gg s$.
Remarkably, $f_{\rm B}$ exhibits a non-monotonic dependence on $\sigma$ when $\nu/s\ll 1$ and $\nu/s\sim 1$, see Fig.~\ref{fig:figure2}~(c,e): the multiplier $f_{\rm B}(\sigma)$ attains a maximum at
 $\sigma_f$ ($0\!<\!\sigma_{f}\!<\!\sigma_{\rm max}$), while
the maximum of $f_{\rm U}$ occurs at   $\sigma_f\approx \sigma_{\rm max}$.
Hence,
the levelling (stabilising) effect of {\rm B}-EN on the competition  compared to {\rm D}-EN  in the slow/intermediate  regimes is maximal for $\sigma\approx \sigma_{f}$. Conversely, $f_{\ell}$ increases with $\sigma$ when $\nu/s\gg 1$, see Fig.~\ref{fig:figure2}~(d,e).

  \begin{figure}[t!]
	\centering
	\includegraphics[width=1\linewidth]{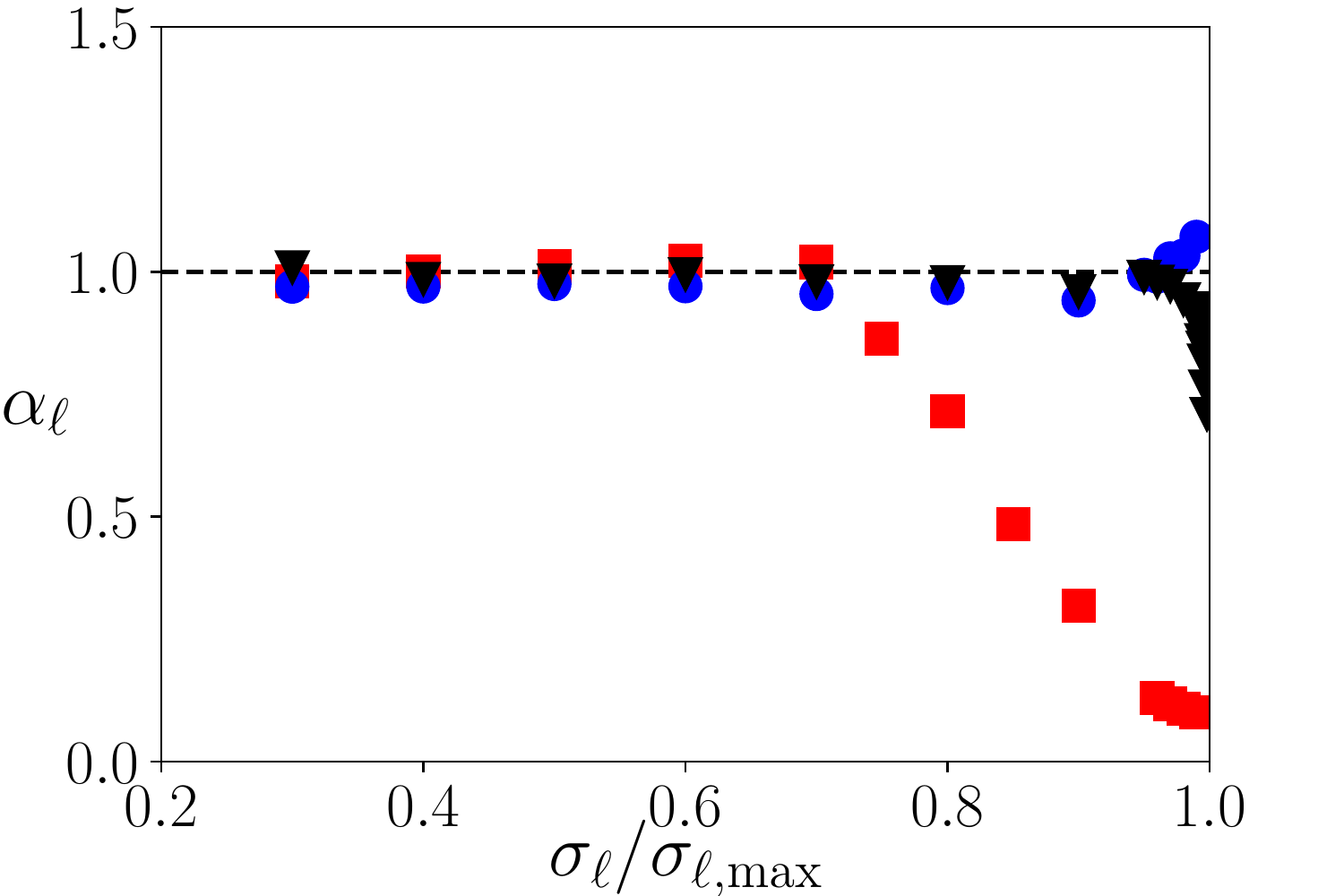}
 \vspace{-4mm}
	\caption{Exponent $\alpha_{\ell}$ of \eqref{scaling}  vs. $\sigma_{\ell}/\sigma_{\ell,\mbox{max}}$ in the regime $\nu/s\gg 1$. Here  $\sigma_{\ell,\mbox{max}}=1/\sqrt{3}$ for
	$\ell\in {\cal L}'$ and $\sigma_{\ell,\mbox{max}}=1$ for ${\rm D}$-EN. Red squares, blue circles and black triangles
	corrsespond to ({\rm B,D,U})-EN, respectively, and dashed line is an eyeguide showing $\alpha_{\ell}=1$. 
	Here $(K_0, s, x_0) = (900, 0.025, 0.5)$, and  different values of $\nu\gg s$,  with $\nu\in[12s,4000s]$~\cite{SM}.
 }
	\label{fig:figure3}
\end{figure}

 Having shown that continuous EN can drastically enhance the fixation probability compared to binary EN,
 we have also studied how  $\phi_{\ell}^\infty$
 is approached when $\nu/s\gg 1$. 
 As $\nu$ increases, we find that $\phi_{\ell}$ converges
 towards
 $\phi_{\ell}^{\infty}$ according to the following scaling, illustrated in Fig.~\ref{fig:figure3}:
\begin{equation}
\label{scaling}
    \phi_{\ell}(\nu) \simeq \phi_{\ell}^{\infty}\exp\left[ {\cal A}_{\ell} (s/\nu)^{\alpha_{\ell}} \right],
\end{equation}
  where the parameter ${\cal A}_{\ell}$ depends on $K_0, \eta$ and $\sigma$.
  In the case of {\rm D}-EN, we found  $\alpha_{\rm D} \approx 1$ regardless of $\sigma$~\cite{TWAM}. 
  In the case of continuous EN,  $\alpha_{\ell \in {\cal L}'}\approx 1$ for small $\sigma$, see Sec.~A2.3 in \cite{SM}, yet $\alpha_{\ell \in {\cal L}'}$  decreases
   as  $\sigma\to \sigma_{\rm max}$,
 indicating a slower convergence to $\phi_{\ell}^\infty$ with ${\rm B/U}$-EN than under ${\rm D}$-EN. As shown in Fig.~\ref{fig:figure3}, the convergence  is particularly slow
   under {\rm B}-EN when $\sigma_{\rm B} \approx \sigma_{\rm max}=1/\sqrt{3}$, while the effect is weaker  under {\rm U}-EN. This stems from $K_{\ell}$ attaining low values with nonzero probability under continuous EN, yielding
   a  slower convergence of  average $N$  to ${
\cal K}_{\ell}$  than under {\rm D}-EN, see Sec.~A.2.2 in \cite{SM}.

We have studied  competition for resources
of two strains  subject  to DN coupled to either binary or continuously-varying EN.
 Our findings suggest that population dynamics is drastically affected by the form of EN:
  continuous EN generally {\it levels the field of competition}
 and significantly increases the fixation probability of the slower strain $S$. This finding is rationalized by mapping results of continuous EN onto those from binary EN, see Fig.~\ref{fig:figure2}. 
 We have also generalized our analysis to a scenario where $S$   produces a public good. Here, we have shown that cooperative behavior greatly benefits from evolving under continuous EN, as  $S$ is typically more likely to fixate than under  binary EN, see Sec. A5 of \cite{SM}.
Our results, demonstrating that discrete and continuous EN, jointly with DN, can have markedly different effects on how populations compete for resources, pave the way to a  better understanding of the influence of such environmental conditions on the evolution of {\it in vivo} microbial communities.

We are grateful to A.M. Rucklidge for useful discussions. A.T. and M.A. acknowledge support from the Israel Science Foundation Grant No. 531/20. M.A. also acknowledges support from the Humboldt Research Fellowship for Experienced Researchers of the Alexander
von Humboldt Foundation. R.W. and M.M. gratefully acknowledge
partial
support of the EPSRC
Ph.D. scholarship EP/N509681/1. M.M. also thankfully acknowledges
partial support of the EPSRC Grant No. EP/V014439/1.
For the purpose of open access, the authors will apply a CC BY public copyright
licence  to any Author Accepted Manuscript version arising.
Data accessibility: supplementary information, simulation source codes and data are electronically available, see Ref.~\cite{SM}.

\onecolumngrid
\newpage
\renewcommand{\theequation}{A\arabic{equation}}
\renewcommand{\thefigure}{S\arabic{figure}}
\setcounter{figure}{0}
\setcounter{equation}{0}
\begin{center}\LARGE{\textbf{Supplemental Material}}
\end{center}
\vspace{4mm}

In this Supplemental Material, we provide some further technical details and supplementary information in support of the results discussed
in the main text. We also provide additional information
concerning the different forms of EN, the
population size distribution, the model and simulation methods, the mean fixation time, and the generalization of the model
to a public good  scenario.

\vspace{0.25cm}
In what follows, unless stated otherwise, the notation is the same as in the main text. Equation~($n$) and Figure~$n$ refer respectively
Equation~($n$) and Figure $n$ of the main text. Reference [$n$] is the item [$n$] of the main text's bibliography. (A$n$) and S$n$ refer to the equation/figure $n$ of this Supplemental Material, respectively.

\vspace{0.25cm}
An e-print of the main text is available at {\bf TBC}.

\vspace{0.25cm}

 Data accessibility: supplementary information, simulations source codes and data are electronically available at https://doi.org/10.6084/m9.figshare.21603480.v1


\section{A1. Different forms of environmental noise}
\label{sec:A1}

For the sake of concreteness and simplicity, our discussion in the main  text focuses mostly on three types of
environmental noise (EN) with symmetric probability density functions (PDFs). Here, we review the main properties of the EN that we have considered in this work.
\subsection{A1.1 Discrete EN}
\label{A1.1}
As a paradigmatic example of discrete EN, we have considered the symmetric \textit{dichotomous} ($\ell={\rm D}$) noise, or ${\rm D}$-EN, $\xi_{\rm D}$ which has been extensively used in the literature, see, {\it e.g.},~[68,69] 
to model a binary switching environment, see, {\it e.g.}, Refs.~[32,33,59,60]
. Here, $\xi_{\rm D} \in \lbrace -\sigma_{\rm D} , \sigma_{\rm D} \rbrace$ is a colored bounded noise whose
stationary PDF is  $p^*(\xi_{\rm D})=\frac{1}{2}~\delta\left(\xi_{\rm D}-\sigma_{\rm D}\right)+
\frac{1}{2}~\delta\left(\xi_{\rm D}+\sigma_{\rm D}\right)$, where $\delta(\cdot)$ denotes the Dirac delta function. The mean, variance and autocorrelation function of the ${\rm D}$-noise at stationarity
are therefore
\begin{eqnarray}
 &&\langle \xi_{\rm D} \rangle \equiv \int_{-\infty}^{\infty}\xi_{\rm D}p^*(\xi_{\rm D})~d\xi_{\rm D}=0,\nonumber\\
 &&{\rm var_D}\equiv\langle \xi_{\rm D}^2 \rangle-\langle \xi_{\rm D} \rangle^2=\sigma_{\rm D}^2,\\
 &&\langle \xi_{\rm D}(t) \langle \xi_{\rm D}(t') \rangle-\langle \xi_{\rm D}(t) \rangle\langle \xi_{\rm D}(t') \rangle=\sigma_{\rm D}^2e^{-\nu|t-t'|}.\nonumber
\end{eqnarray}


The properties of ${\rm D}$-EN can  readily be generalized to the case of asymmetric switching, see, e.g., Refs.~[59,68,69].

\subsection{A1.2 Continuous EN}
\label{A1.2}
We have considered two paradigmatic examples of continuous EN, $\xi_{\ell}$ with $\ell\in {\cal L}'=\{{\rm B}, {\rm U}\}$, corresponding to beta and uniform EN, respectively. Both the beta and uniform distributions are used in a wide variety of scientific disciplines and are suitable to describe the behavior of random variables of finite interval length with zero and finite lower bounds, respectively. Below we outline  the derivation of the  diffusive term ${\cal B}_{\ell}$ given by Eq.(4) in the main text, and then review the main statistical properties of $\xi_{\ell}$ and those of the carrying capacity $K_{\ell}$.

\subsubsection{A1.2.1 Derivation of the diffusion term ${\cal B}_{\ell}$ and the stationary PDF}
\label{A1.2.1}
The continuous noise $\xi_{\ell}$ is generally defined by
 Eq. (3) in the main text on the domain  $[c_{\ell}, d_{\ell}]$, and the stationary PDF is therefore the solution of
\begin{eqnarray}
 \label{eq:zero-cur}
 0=\left[\nu\xi_{\ell}+\frac{1}{2}\frac{\partial}{\partial \xi_{\ell}}{\cal B}_{\ell}\right]p^*(\xi_{\ell}),
\end{eqnarray}
which ensures that the current of probability vanishes at $\xi_{\ell}=c_{\ell},d_{\ell}$~[28,29,67].

In order to derive the diffusion term  corresponding to a specified
stationary PDF $p^*(\xi_{\ell})$, we integrate \eqref{eq:zero-cur}
and get the following closed expression satisfied by   ${\cal B}_{\ell}$:
\begin{eqnarray}
 \label{eq:B-derivation}
 {\cal B}_{\ell}(\xi_{\ell}) = {\cal B}_{\ell}(c_{\ell}) -2\nu \frac{\int_{c_{\ell}}^{\xi_{\ell}}\xi_{\ell} p^*(\xi_{\ell})d\xi_{\ell}}{p^*(\xi_{\ell})}.
\end{eqnarray}
Clearly ${\cal B}_{\ell}(c_{\ell})={\cal B}_{\ell}(d_{\ell})$ if the stationary average of
$\xi_{\ell}$  vanishes, i.e. $\langle \xi_{\ell} \rangle \equiv \int_{c_{\ell}}^{d_{\ell}}\xi_{\ell}p^*(\xi_{\ell})~d\xi_{\ell}=0$. This is the case here since  we focus on symmetric EN. Moreover, we choose the domain
$[c_{\ell},d_{\ell}]$ to be symmetric, $d_{\ell}=-c_{\ell}$, and such that ${\cal B}_{\ell}(c_{\ell})={\cal B}_{\ell}(d_{\ell})={\cal B}_{\ell}(-c_{\ell})=0$, which ensures that the  diffusion term vanishes at the boundaries $\xi_{\ell}=c_{\ell},d_{\ell}$. Hence, the diffusion term ${\cal B}_{\ell}$ for the $\ell\in\{{\rm U}, {\rm B}\}$-EN considered in the main text is
\begin{align}
 \label{eq:Bell}
 {\cal B}_{\ell}(\xi_{\ell}) =  -2\nu \frac{\int_{c_{\ell}}^{\xi_{\ell}}\xi_{\ell} p^*(\xi_{\ell})d\xi_{\ell}}{p^*(\xi_{\ell})}
\end{align}
for $\ell\in \{{\rm B}, {\rm U}\}$. The explicit expressions of ${\cal B}_{\rm B}$ and ${\cal B}_{\rm U}$ are given by Eq. (4) in the main text.

The complementary equation to \eqref{eq:B-derivation} giving the expression of the stationary PDF  $p^*(\xi_{\ell})$ in terms of ${\cal B}_{\ell}$, is obtained by solving Eq. \eqref{eq:zero-cur}
\begin{eqnarray}
 \label{eq:p-def}
 p^*(\xi_{\ell})= \frac{{\cal N}}{{\cal B}_{\ell}(\xi_{\ell})}~{\rm exp}\left(-2\nu\int_{c_{\ell}}^{\xi_{\ell}}\frac{\xi_{\ell}}{{\cal B}_{\ell}(\xi_{\ell})}~d\xi_{\ell}\right),
\end{eqnarray}
with the normalization \[{\cal N}^{-1}\equiv \int_{c_{\ell}}^{d_{\ell}}
\frac{d\xi_{\ell}}{{\cal B}_{\ell}(\xi_{\ell})}~{\rm exp}\left(-2\nu\int_{c_{\ell}}^{\xi_{\ell}}\frac{\xi_{\ell}}{{\cal B}_{\ell}(\xi_{\ell})}~d\xi_{\ell}\right).\]
\subsubsection{A1.2.2 Properties of the different forms of continuous EN}
\label{A1.2.2}
We now review the main properties of the stationary continuous EN in the case of  beta and uniform noise.
Since the varying carrying capacity is here given by Eq.(2):
\begin{align}\label{eq:Kl}K_\ell(t) =K_{0}[1+\xi_{\ell}(t)],\end{align}
we readily obtain the stationary PDF of the time-varying carraying capacity:
\begin{eqnarray}
 \label{eq:zero-cur2}
 {\cal P}(K_{\ell})&\equiv&
 \frac{1}{K_{0}}p^*\left(\frac{K_{\ell}}{K_{0}}-1\right),
\end{eqnarray}
whose support is ${\cal S}_{\ell}=[K_{0}(1+c_{\ell}), K_{0}(1+d_{\ell})]=[K_{0}(1-d_{\ell}), K_{0}(1+d_{\ell})]$.

We can now consider explicitly the different forms of continuous EN
used in this work:
\begin{enumerate}
  \item[-] Symmetric beta noise ($\ell={\rm B}$, {\rm B}-EN):
  we have considered the single-parameter ($\beta>1$)
  colored symmetric continuous noise $\xi_{\rm B}$ whose stationary PDF
  is the beta distribution obtained from Eqs.~\eqref{eq:zero-cur} and (4) of the main text,
  with $c_{\rm B}=-1$ and $d_{\rm B}=1$, which yields
  $p^*(\xi_{\rm B})=\left[2^{1-2\beta}/{\rm B}\left(\beta,\beta\right)\right]\left(1+\xi\right)^{\beta-1}\left(1-\xi\right)^{\beta-1}$, where $B\left(\beta,\beta\right)\equiv\int_{0}^{1} t^{\beta-1}(1-t)^{(\beta-1)} ~dt$ is the usual beta function.

  We thus find the moments of $\xi_{\rm B}$:
  \begin{eqnarray}
 &&\langle \xi_{\rm B} \rangle \equiv \int_{-1}^{1}\xi_{\rm B}p^*(\xi_{\rm B})~d\xi_{\rm B}=0,\nonumber\\
 &&{\rm var_B}\equiv\langle \xi_{\rm B}^2 \rangle-\langle \xi_{\rm B} \rangle^2=\sigma_{\rm B}^2=1/(2\beta +1)<\sigma_{\rm max}^2,\quad \text{ where $\sigma_{\rm max}\equiv 1/\sqrt{3}$.}
\end{eqnarray}

   \item[-] Symmetric uniform noise ($\ell={\rm U}$, {\rm U}-EN):
  we have also considered the single-parameter ($0<\sigma_{\rm U}<\sigma_{\rm max}$)
  symmetric continuous noise $\xi_{\rm U}$ whose stationary PDF
  is the uniform distribution obtained from Eqs.~\eqref{eq:zero-cur} and (4) of the main text,
  with $c_{\rm U}=-\sqrt{3}\sigma_{\rm U}$ and $d_{\rm U}=\sqrt{3}\sigma_{\rm U}$, which yields
  $p^*(\xi_{\rm U})=1/\left(2\sqrt{3}\sigma_{\rm U}\right)$.

  We thus find the moments of $\xi_{\rm U}$:
  \begin{eqnarray}
 &&\langle \xi_{\rm U} \rangle \equiv \int_{-\sqrt{3}\sigma_{\rm U}}^{\sqrt{3}\sigma_{\rm U}}\xi_{\rm U}p^*(\xi_{\rm U})~d\xi_{\rm U}=0,\nonumber\\
 &&{\rm var_U}\equiv\langle \xi_{\rm U}^2 \rangle-\langle \xi_{\rm U} \rangle^2=\sigma_{\rm U}^2<\sigma_{\rm max}^2.
\end{eqnarray}
Note that in order to ensure that $K_{{\rm U}}>0$, we require $c_{\rm U}>-1$ and thus $\sigma_{\rm U}<\sigma_{\rm max}$.

\end{enumerate}

In addition to the beta and uniform noise, we can consider  other examples of continuous EN. A notable example is the Ornstein-Uhlenbeck ($\ell={\rm OU}$) noise with Gaussian statistics. In fact, since  OU noise is unbounded, one must consider a version of this noise with a truncated range
to ensure that carrying capacity never becomes negative. Yet,  the OU truncation yields cumbersome expressions for the noise moments, and more importantly, it is found to lead to results that are highly sensitive to the details of
the truncation procedure at $K_{{\rm OU}}\to 0$. In light of these considerations, we  have chosen not to use further the {\rm OU}-EN in this study.

Notably, we have checked  that our main results here and in the main text also hold for different forms of \textit{asymmetric EN}, like asymmetric dichotomous noise and EN obeying an asymmetric Gamma distribution (Gamma EN). In particular, we have confirmed that for Gamma EN a similar scaling behavior reported in Fig.~3 of the main text (for {\rm B}-EN), with a decreasing  exponent occurs, when $\sigma$ is increased, see details in the main text.

\section{A2. Population size distribution}
\label{sec:SM2}
The population size distribution (PSD)
plays a key role in this work as, e.g., it enters into  Eq. (5) of the main text. To appreciate the fundamental differences in modelling population dynamics subject to discrete or continuous stationary EN, it is useful to consider first the master equation (ME) for the PSD, $P(N,t)|_{K_0}$, in a \textit{static} environment with a constant carrying capacity $K\equiv K_0$. For the birth-death process associated with the rates  $T_{S/F}^{+}=(f_{S/F}/\bar{f})~N_{S/F}$ and
$T_{S/F}^{-}=(N/K_0)~N_{S/F}$, and with $N=N_S+N_F$, $\;T_{S}^{+}+T_{F}^{+}=N$ and $T_{S}^{-}+T_{F}^{-}=N^2/K_0$, the ME reads
 \begin{equation}
  \label{eq:ME1}
\frac{\partial P(N, t)}{\partial t}\Big|_{K_0}
 = (\mathbb{E}_N^{-}-1)[N P(N,t)|_{K_0}] + \frac{(\mathbb{E}_N^{+}-1)}{K_0}\left[N^2~P(N,t)|_{K_0}\right],
 \end{equation}
 where  $\mathbb{E}_N^{\pm}$ are shift operators such that
 $\mathbb{E}_N^{\pm} f(N,t)=
 f(N\pm 1,t)$ for any suitable $f(N,t)$.
 The long-lived PSD of this single-variate ME is found by setting $\dot{P}(N, t)|_{K_0}=0$
 in \eqref{eq:ME1} and by imposing
 a reflecting boundary condition (BC) at $N=1$, yielding  $P(N)|_{K_0}\simeq \frac{K_0}{N}\frac{K_0^{N}e^{-K_0}}{N!}$
 when $K_0\gg1$~[59].

\begin{figure}
	\centering
	\includegraphics[width=0.61\linewidth]{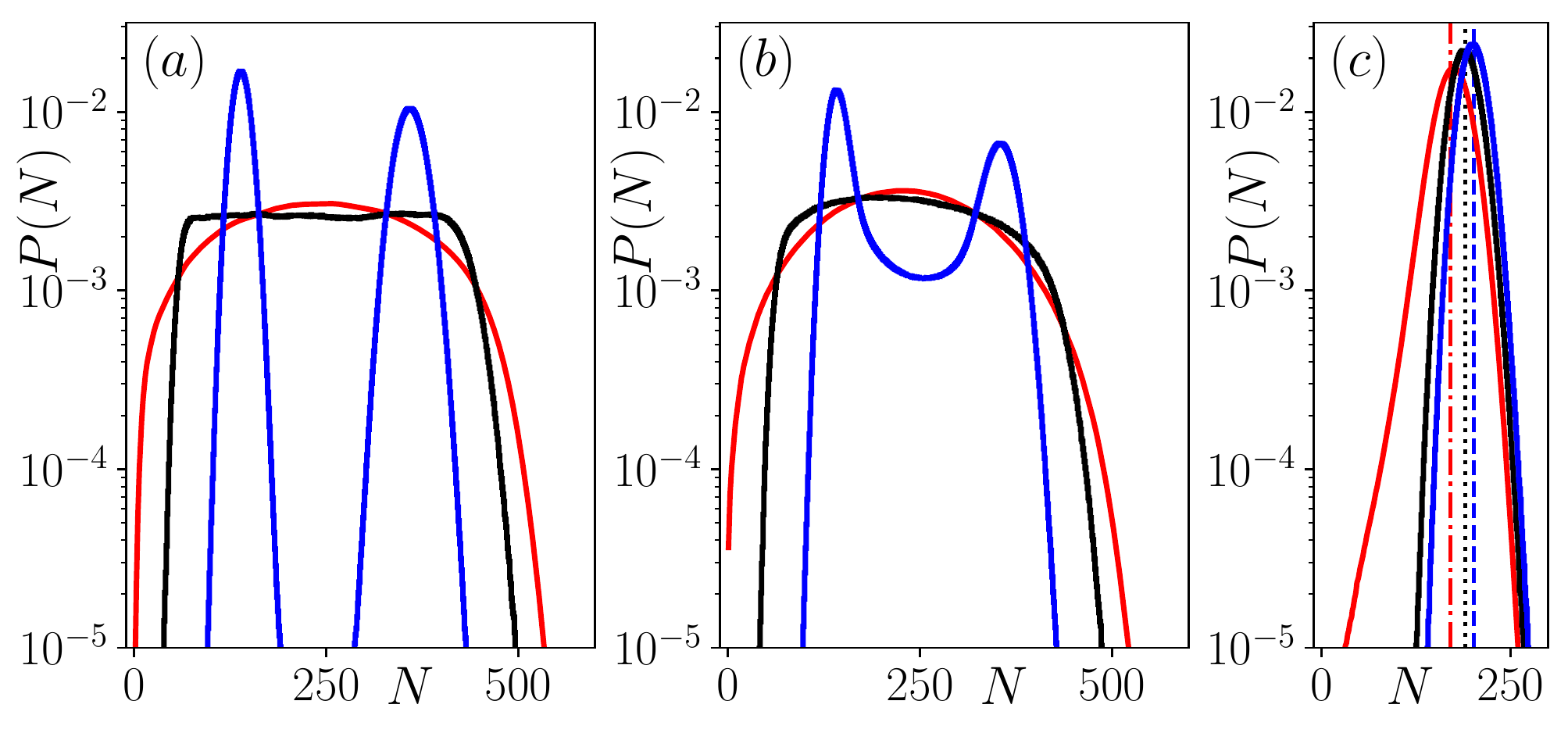}	\includegraphics[width=0.61\linewidth]{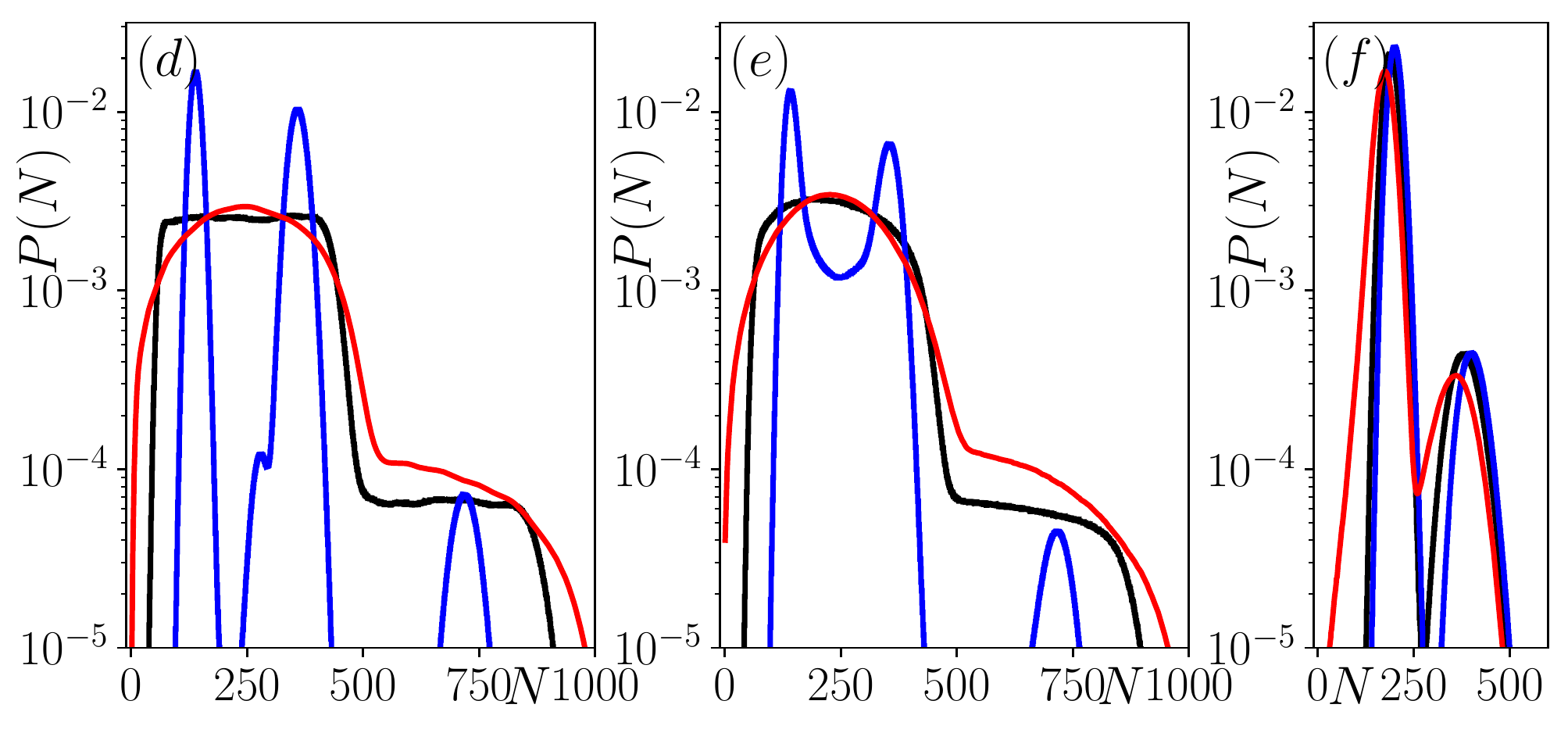}
	\vspace{-3mm}
    \caption{Histograms of the quasi-stationary PSD $P_{\ell}(N,\nu)$  for $\ell={\rm B}$ (beta EN, red / gray), $\ell={\rm U}$ (uniform EN, black) and $\ell={\rm D}$ (dichotomous EN, blue / dark gray) EN, with $\nu = (10^{-3}, 10^{-0.5}, 10^{2})$ in (a,d), (b,e), (c,f) respectively and $b=0$ in (a,b,c) and $b=1$
	 in (d,e,f), see Sec.~A5.
	 To ensure that PSD is at quasi-stationarity, we have sampled the realizations after $t_0=16/s$.
	 Other parameters are $(K_0,\sigma_{\ell}=\sigma,s,x_0) = (250,0.44,0.02,0.5)$. When $b=0$, all forms of noise lead to unimodal distributions when $\nu$ is large, but have different forms at small and intermediate $\nu$: $P_{\rm D}$ is dual peaked, $P_{\rm B}$ is unimodal, while $P_{\rm U}$ has a  flat shape for small $\nu$ that becomes unimodal for intermediate and larger $\nu$. This, in turn, leads to different functional forms when $b>0$: $P_{\rm D}$ can have $2, 3$ or $4$ peaks (depending on $b$ and $\nu$, see [32,33]), 
	 $P_{\rm B}$ is unimodal with a ``shoulder'' at low and intermediate $\nu$ (d,e) and bimodal for large $\nu$ (f), while $P_{\rm U}$ has a two-step shape for small $\nu$ (d), transformed into two asymmetric ``shoulders'' at intermediate $\nu$ (e) that become two sharp peaks at large $\nu$ (f).
	  }
	\label{fig:pdfs}
\end{figure}
\subsubsection{A2.1 Properties of $P_{\ell}(N,\nu)$ and  mean population size $\langle N \rangle_\ell$  in a changing environment}
\label{A2.1}

In a changing environment, the carrying capacity varies according to Eq. (2) of the main text, where  $\xi_{\ell}$ denotes the EN that can be either discrete ($\ell={\rm D}$) or continuous ($\ell\in \{{\rm B, U}\}$). The PSD, $P(N,\xi_{\ell}, t)$ thus obeys \eqref{eq:ME1}, with $K_0 \to K_{\ell}(t)$ given by Eq. (2) of the main text, which results  in a coupling of the
ME with the EN dynamics.

When EN is discrete, this coupling yields a multivariate ME. In the case of symmetric dichotomous noise $\xi_{{\rm D}}\in \{-\sigma_{\rm D}, \sigma_{\rm D}\}$ which switches,
according to $\xi_{{\rm D}} \to -\xi_{{\rm D}}$, at rate $\nu/2$, the resulting ME is
  \begin{equation}
  \label{eq:ME2}
\frac{\partial P(N, \xi_{\rm D}, t)}{\partial t}=
  (\mathbb{E}_N^{-}-1)[N P(N,\xi_{\rm D},t)] + \frac{(\mathbb{E}_N^{+}-1)}{K_{0}[1+\xi_{{\rm D}}(t)]}\left[N^2~P(N,\xi_{\rm D}, t)\right]+ \frac{\nu}{2} \left[P(N,-\xi_{\rm D},t) -P(N,\xi_{\rm D},t)\right],
 \end{equation}
 where the last line encodes binary environmental switching at rate $\nu/2$. The dynamics underpinned by this ME
 can be simulated {\it exactly} with the standard Gillespie algorithm~[75] 
 or a variant of it~[76], 
 see Sec.~A.3. Similar multivariate MEs can be obtained for more general forms of discrete noise, {\it e.g.} asymmetric binary switching~[59] 
 or random switching in a randomly cyclic changing environment~[60]. 
 In the case of dichotomous noise, the birth-death process described by \eqref{eq:ME2} can be aptly approximated by a suitable {\it piecewise deterministic process} (PDMP)~[73]. 
 In the realm of the PDMP, which ignores entirely demographic noise (DN),
  the  (marginal) PSD, here  denoted by $P_{\rm D}(N,\nu)=\sum_{\xi_{\rm D}=\pm \sigma_{\rm D}}P(N,\xi_{\rm D})$, is approximated by~[32,33,59]
\begin{equation}
\label{PSD_PDMP}
P_{\rm D}(N,\nu)\simeq
P_{\rm D}^{\rm PDMP}(N,\nu)\propto\frac{1}{N^2}\left[\frac{((1+\sigma_{\rm D})K_{0} -N)(N-(1-\sigma_{\rm D})K_{0})}{N^2}\right]^{\frac{\nu}{2}-1},
\end{equation}
where  we have omitted the normalization constant.  Equation~\eqref{PSD_PDMP} provides a good approximation of the PSD when  $K_{0}\gg 1$ and
  $\nu\ll K_{0}$, i.e., when EN dominates over DN, and allows for an accurate approximation of $\phi_{\rm D}$ using Eq. (5) of the main text. 

When EN is continuous,  the ME \eqref{eq:ME2} with $\xi_{\rm D}$ substituted by $\xi_{\ell}$, is coupled with the stochastic differential equation (SDE)~(3) of the main text (interpreted as It\^o SDE~[67]). 
The {\it discrete} birth-death dynamics is hence coupled to  $\xi_{\ell}$ that varies continuously with a  correlation time $1/\nu$.
 In this way, the coupled process does not admit a unique interpretation and is therefore generally non-Markovian.

In order to work with only discrete random variables, and to restore the Markov property, we interpret the coupling of (1)-(3) by using the well-known relationship between It\^o SDEs and the continuous limit of ME (assuming a large carrying capacity) resulting in a {\it fictitious birth-death process}, that is a discrete counterpart of the
 SDE~(3) of the main text~[77].
The ME for the long-time PSD $P\left(N,K_\ell,t\right)$ is given by:

  \begin{eqnarray}
  \label{eq:ME-full}
\hspace{-3cm}&&\frac{\partial P(N, K_{\rm \ell}, t)}{\partial t}=
  (\mathbb{E}_N^{-}\!-\!1)[N P(N,K_{\rm \ell},t)] + \frac{(\mathbb{E}_N^{+}\!-\!1)}{K_{\ell}(t)]}\left[N^2~P(N,K_{\rm \ell}, t)\right]\nonumber \\
  &&+(\mathbb{E}_{K_\ell}^{-}\!-\!1)\left\{\frac{1}{2}\left[K_0^2  {\cal B}_{\ell}-\nu\left(K_{\ell}-K_0\right)\right] P(N,K_{\rm \ell},t)\right\}+ (\mathbb{E}_{K_\ell}^{+}\!-\!1)\left\{\frac{1}{2}\left[K_0^2  {\cal B}_{\ell}+\nu\left(K_{\ell}-K_0\right)\right] P(N,K_{\rm \ell},t)\right\},
 \end{eqnarray}
 %
	%
	where ${\cal B}_\ell$, given by (4) and \eqref{eq:Bell}, is evaluated at $\xi_\ell=\frac{K_\ell}{K_0}-1$, and $\mathbb{E}_{K_\ell}^{\pm}$ are shift operators acting on the carrying capacity $K_\ell$ such that
 $\mathbb{E}_{K_\ell}^{\pm} f(N,K_\ell,t)=
 f(N,K_\ell\pm 1,t)$ for any suitable function $f(N,K_\ell,t)$ of $N$ and $K_\ell$ (similarly, $\mathbb{E}_N^{\pm}$ are shift operators acting on $N$).
 In a suitable continuous limit, the environmental dynamics is then described by the SDE (3) while the evolutionary dynamics is encoded in the Fokker-Planck equation  associated to the continuous limit (diffusion approximation) of the first line of \eqref{eq:ME-full}.
  Here,  the marginal PSD in the presence of continuous EN is then $
  P_{\ell}(N,\nu)=\int_{K_{0}(1+c_{\ell})}^{K_{0}(1+d_{\ell})} P(N,K_{\rm \ell})~dK_\ell$.
  Analytical progress 
  is possible in the limits $\nu \to 0$ (adiabatic noise), and $\nu\to\infty$ (short-correlated noise), see main text and Sec. A3 for the practical simulation of this process.
%

General properties of the system, like fixation probability and mean extinction time,  depend greatly on the statistics of $P_{\ell}(N,\nu)$. The similarities and differences between these properties under the influence of different types of EN can in part be explained by the properties of  $P_{\ell}(N,\nu)$.

The PSD under ${\rm D}$-noise has been studied in Refs.~[32,33,59] 
where it was shown to be characterized by a noise-induced transition arising at $\nu=2$:
$P_{\rm D}(N,\nu)$ moves from being bimodal ($\nu<2$) to unimodal ($\nu>2$) as $\nu$ is increased,
and is generally well approximated by \eqref{PSD_PDMP}~[32,33,59-61]. 
In stark contrast, $P_{\rm B}(N,\nu)$
and $P_{\rm U}(N,\nu)$ undergo {\it no noise-induced transition}   when $\nu$ varies: as illustrated in  Fig.~\ref{fig:pdfs}(a,b,c), these continuous noise PSD are essentially unimodal (or flat)
{\it for all values of $\nu$}.  In fact, when  $\nu \ll 1$, the PSD follows the variations of $K_{\ell}(t)$
  and hence $P_{\ell} (N,0)\approx {\cal P}(K_{\ell})$, given by Eq.~\eqref{eq:zero-cur2}, which is bimodal for ${\rm D}$-EN, unimodal for  ${\rm B}$-EN, and flat for ${\rm U}$-EN, as reported in \ref{fig:pdfs}(a). When  $1\ll \nu \lesssim K_{0}$, the effect of EN dominates over DN, and
  the PSD is unimodal, see Fig.~\ref{fig:pdfs}(c),
  with $N\approx {\cal K}_{\ell}$ when $\sigma_{\ell} \ll 1$ and not too close to $\sigma_{\rm max}$ [${\cal K}_{\ell}$ is given by Eq.~(8) in the main text, see also below].  For intermediate values of $\nu$, when $\nu$ is increased, $P_{\rm D}$ morphs from being bimodal to gradually becoming unimodal, while $P_{\rm B}$
and $P_{\rm U}$  globally retain their shapes that become slightly narrower, compare Figs.~\ref{fig:pdfs}(a,b).
  Notably, a more complex picture for the PSD emerges in a scenario with public good production by the $S$ strain, see Sec.~A5 and Fig.~\ref{fig:pdfs}(d-f).

  \vspace{0.25cm}
  \begin{figure}
	\centering
	\includegraphics[width=0.4\linewidth]{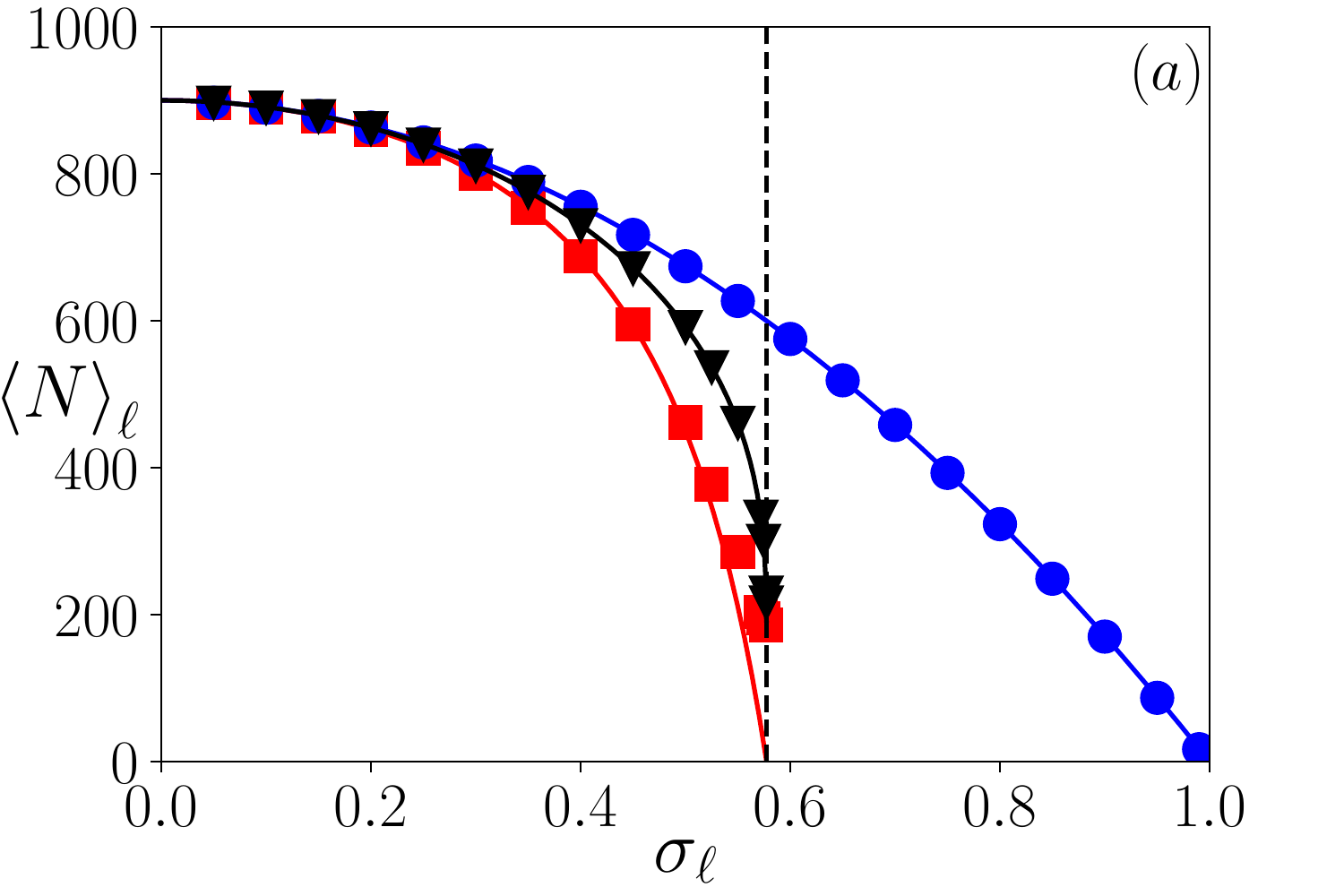}
	\includegraphics[width=0.4\linewidth]{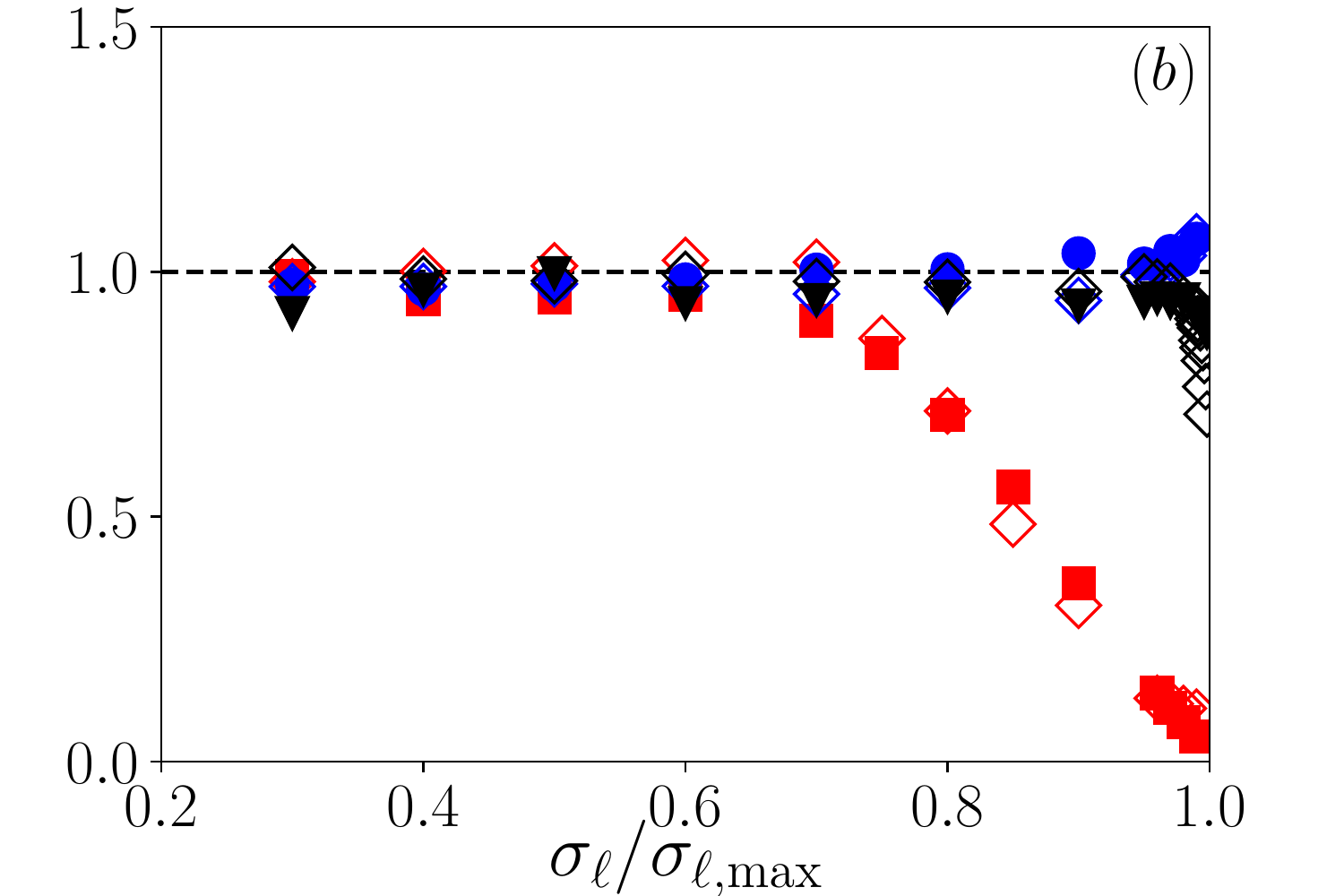}

	\vspace{-4mm}\caption{(a)  $\langle N \rangle_\ell$ (mean of the PSD) vs. $\sigma_{\ell}$ for $\nu\gg 1$ ($\nu=1000$).
	Simulation results for {\rm D}-EN, {\rm U}-EN, and {\rm B}-EN are represented by blue circles, black triangles and red squares respectively. Solid lines show ${\cal K}_{\rm D}$ (blue),  ${\cal K}_{\rm U}$ (black) and ${\cal K}_{\rm B}$ (red) given by Eq.(8) in the main text. Black dashed line show the value of   $\sigma_{\ell,\mbox{max}}=1/\sqrt{3}$ for the continuous EN.
	(b) Exponents $\theta_{\ell}$ of $(s/\nu)$ for $\ln \left(\langle N\rangle_{\ell}/\lim_{\nu \to \infty}\langle N\rangle_{\ell}\right)$ and $\alpha_{\ell}$ of $(s/\nu)$ for $\ln \left(\phi_{\ell}/\phi_{\ell}^{\infty}\right) $ vs. $\sigma_{\ell}/\sigma_{\ell,{\rm max}}$
	in the regime of high varying  rate $\nu$. Here, $\theta_{\ell}$ and $\alpha_{\ell}$ are respectively represented by full markers and hollow diamonds markers, and have been obtained for different values of $\nu$ (with $\nu/s\gg 1$), as in Fig.~3. Note that $\sigma_{\ell,\mbox{max}}=1$ for {\rm D}-EN. We find the exponents $\theta_{\rm D}\approx \alpha_{\rm D}\approx 1$ for {\rm D}-noise at all $0<\sigma_{\rm D}<1$. For {\rm B}- and {\rm U}-EN the exponents still (approximately) coincide,  $\theta_{\rm B}\approx \alpha_{\rm B}$ and $\theta_{\rm U}\approx \alpha_{\rm U}$,
	but are approximately equal to 1 only  when $\sigma_\ell \ll\sigma_{\ell,{\rm max}} $, while they are less than 1 as $\sigma_\ell$ approaches $\sigma_{\ell,{\rm max}}$.
  Parameters are $(K_0, s, x_0) = (900, 0.025, 0.5)$.
	}
	\label{fig:moments}
\end{figure}

In Fig.~\ref{fig:moments}(a), we show the long-time population size average given by
\begin{equation}
\langle N \rangle_\ell\equiv \int_0^{\infty} N P_{\ell}(N,\nu)~dN,
\end{equation}
as a function of $\sigma_{\ell}$ in the regime of high varying rate ($\nu \gg 1$). We find that $\langle N \rangle_\ell$ essentially coincides with ${\cal K}_{\ell}$ as long as $\sigma_{\ell}$ is not too close to $\sigma_{{\ell},{\rm max}}$ ; here $\sigma_{{\ell},{\rm max}}=1/\sqrt{3}$ for $\ell= {\rm B}, {\rm U}$, and $\sigma_{{\ell},{\rm max}}=1$ for $\ell= {\rm D}$. Deviations arise when $\sigma_{\ell}$ approaches $\sigma_{{\ell},{\rm max}}$, which are particularly visible for ${\rm B}$-EN.
We have found that these deviations can be captured by realizing that in regime of high varying rate, as
$\sigma_{\ell}$ approaches $\sigma_{{\ell},{\rm max}}$, the mean population size scales as  $\langle N\rangle_\ell \sim {\cal K}_{\ell}~e^{(s/\nu)^{\theta_{\ell}}}$. The dependence of $\theta_{\ell}$ on $\sigma_{\ell}$ is shown in Fig.~\ref{fig:moments}(b), where it is found to match the  dependence of the exponent $\alpha_{\ell}$ defined by Eq.~(9) in the main text, and governing the convergence of   $\phi_{\ell}\to \phi_{\ell}^{\infty}$, as illustrated by  Fig.~3.
This can be explained by the fact that
the convergence  $\phi_{\ell}\to \phi_{\ell}^{\infty}$ is highly dependent on the mean of $N$ that can vary greatly for continuous EN
when their variance approaches $\sigma_{{\ell},{\rm max}}$.
This suggests $\phi_{\ell}\approx  {\rm exp}(-s \langle N\rangle_\ell)$ when $\nu/s\gg 1$. With the scaling of Eq.~(9),  this leads to the results reported in Figs.~3 and \ref{fig:moments}(b) with $\alpha_{\ell}\approx \theta_{\ell}$.


  %
\subsubsection{A2.2 Derivation of ${\cal K}_{\ell}$}
\label{A2.2}

\begin{figure}
	\centering
	\includegraphics[width=0.45\linewidth]{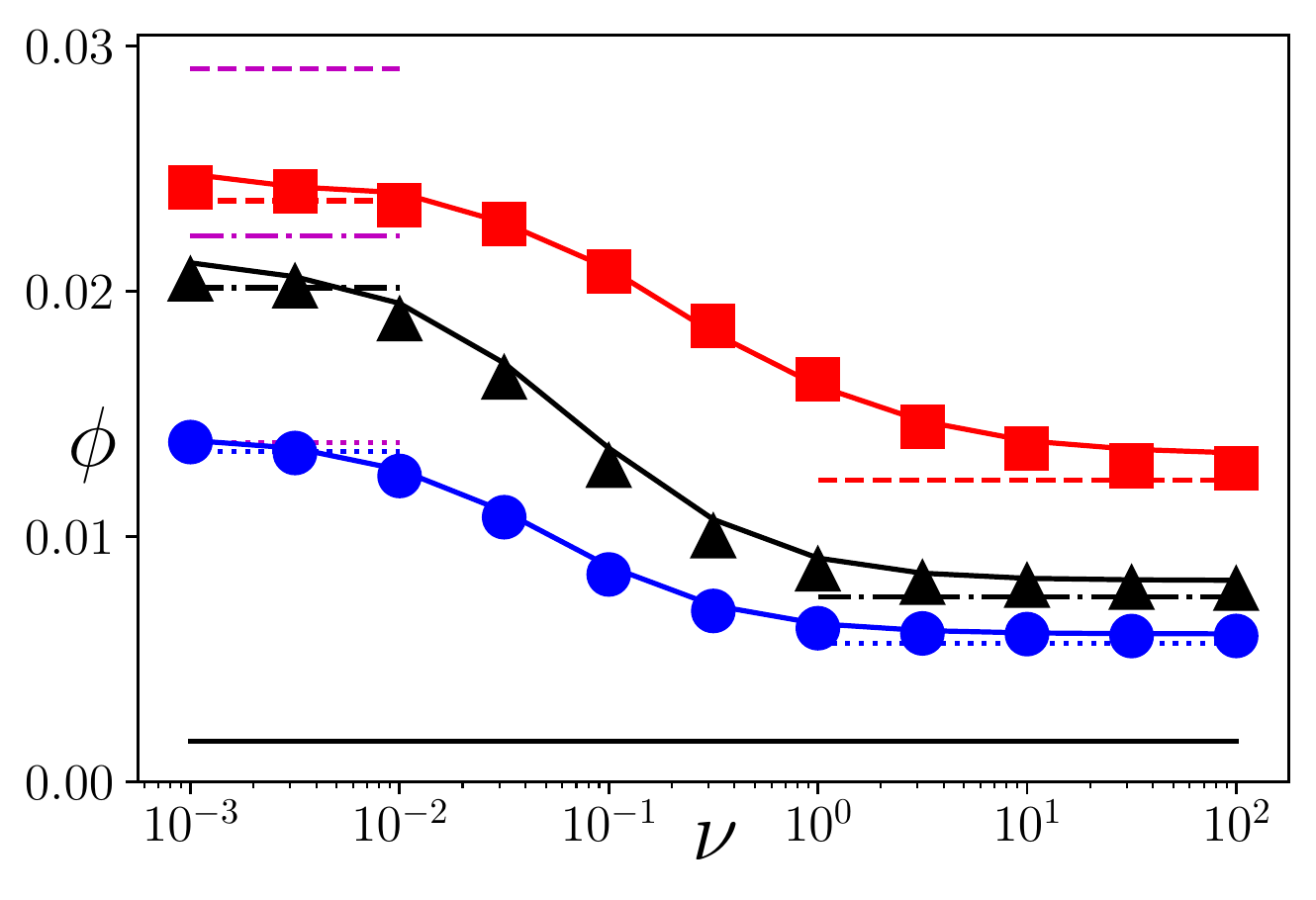}	
	\vspace{-5mm}\caption{ $S$-fixation probability $\phi_{\ell}$ under $\ell$-EN vs. $\nu$ over many decades. Symbols are from simulation data (with the fictitious chemical reaction method, see Sec. A3),
	and solid lines (red / gray: $\ell={\rm B}$, black: $\ell={\rm U}$, blue / dark gray: $\ell={\rm D}$) are from numerical evaluation of Eq.~(5) of the main text using histograms obtained from simulation data (see Fig.~\ref{fig:pdfs} and Sec. A3). Dashed ($\ell={\rm B}$), dashed-dotted ($\ell={\rm U}$) and dotted ($\ell={\rm D}$) lines show $\phi_{\ell}^{(0,\infty)}$. $\phi_{\ell}^{0}$ is obtained from the numerical evaluation of Eq. (5) using analytical results for $P_{\ell}$ under low $\nu$ (red/black/blue lines) and Eq.~(6) (magenta lines), while $\phi_{\ell}^{\infty}$ is given by Eq.~(8), using $\phi_{\ell}^\infty\approx\exp\left(-\eta\mathcal{K}_{\ell}\right)$.
	For all $\ell$-EN, simulation data and predictions of  Eq.~(5) are in good agreement with
	$\phi_{\ell}^{\infty}$ when $\nu/s\gg 1$, and with  $\phi_{\ell}^{0}$ when $\nu/s\ll 1$.
	The black solid horizontal line shows $\phi(K_0, s, x_0)$, the $S$ fixation probability in a static environment. Clearly $\phi_{\ell}(\nu)\gg\phi(K_0, s, x_0)$ over then entire range of $\nu$ for $\ell\in\{{\rm B},{\rm D}, {\rm U}\}$, and $\phi_{\rm B}(\nu) > \phi_{\rm U}(\nu) > \phi_{\rm D}(\nu)$.
	Parameters are: $(K_0,\sigma_{\ell}=\sigma, s,x_0)=(250,0.44,0.05,0.5)$. 
	%
	}

	\label{fig:newFigS3}
\end{figure}

 Since ${\cal K}_{\ell}$ plays an important role in our analysis when $\nu\gg 1$, it is useful to outline its derivation. When the population size is large and $\nu\gg 1$, the stochastic logistic equation $\frac{d}{dt}N(t)=N(t)(1-N(t)/K_{\ell}(t))$
is well described by replacing $1/K_{\ell}(t)$ by its average over the stationary EN probability density, $\langle 1/K_{\ell}(t)\rangle$. This stems from the self averaging of  EN, resulting from the many switches occurring prior $N$ settles in its PSD~[32,33,59], 
yielding
\begin{equation}
    \frac{d}{dt}N(t)=N(t)\left(1-N(t)\left\langle \frac{1}{K_{\ell}(t)}\right\rangle \right)\equiv N(t)\left(1-\frac{N(t)}{{\cal K}_{\ell}} \right),
\end{equation}
where, with \eqref{eq:Kl}, we have
\begin{align}
\label{eq:Kell}
 \frac{1}{{\cal K}_{\ell}}\equiv \left\langle \frac{1}{K_{\ell}(t)}\right\rangle=\frac{1}{K_0}\left\langle \frac{1}{1+\xi_{\ell}(t)}\right\rangle\equiv
 \frac{1}{K_0}\int_{-d_{\ell}}^{d_{\ell}}\frac{p^*(\xi_{\ell})}{1+\xi_{\ell}}\, d\xi_{\ell}.
\end{align}
${\cal K}_{\ell}$ can thus be computed directly from $p^*(\xi_{\ell})$ given in Sec.~A1.2, yielding the expressions given by Eq.(8) in the main text. As an example, we consider the case of symmetric dichotomous noise ($\ell={\rm D}$): at stationarity, $\xi_{\rm D}=\pm \sigma_{\rm D}$ with probability $1/2$, and \eqref{eq:Kell} thus boils down to:
${\cal K}_{\rm D}^{-1}=(1/2)K_0^{-1}\left[(1+\sigma_{\rm D})^{-1}+(1-\sigma_{\rm D})^{-1}\right]=[K_0(1-\sigma_D^2)]^{-1}$.
Hence, for symmetric {\rm D}-EN, we obtain ${\cal K}_{\rm D}=K_0(1-\sigma_{\rm D}^2)$, which is the harmonic mean of $K_0(1-\sigma_{\rm D})$ and $K_0(1+\sigma_{\rm D})$, the two possible values of $K_{\rm D}$~[32,33]. 
In  Eq.(8), we report the expressions ${\cal K}_{\ell}$ that are such that
${\cal K}_{\rm B} \leq {\cal K}_{\rm U}\leq {\cal K}_{\rm D}$, see Figs.~\ref{fig:pdfs}(c) and \ref{fig:moments}(a). Since
$\phi_{\ell}^{\infty}\approx    {\rm exp}(-s{\cal K}_{\ell})$
when $\nu/s\gg 1$, this readily yields $\phi_{\rm B}^{\infty} \geq \phi_{\rm U}^{\infty} \geq \phi_{\rm D}^{\infty}$ in the regime of high varying rate, as confirmed in Fig.~\ref{fig:newFigS3}. This figure also
and corroborates the results reported in the main text
according to which $\phi_{\rm B}(\nu) \geq \phi_{\rm U}(\nu) \geq \phi_{\rm D}(\nu)\gg\phi(K_0, s, x_0)$.

\subsubsection{A2.3 WKB approximation}
\label{A2.3}
In this section, we calculate the PDF of $N$, and the fixation probability $\phi_{\ell}$ of the slow growers in the regime of high varying rate ($\nu/s\gg 1$) using the WKB approximation. While we demonstrate our method with the beta-distributed EN ({\rm B}-EN), it can be shown that our results are generic, and can be generalized to
any EN distribution whose statistics  about the mean can be well approximated by  a suitable Gaussian (in the regime of short-correlated EN), which requires the variance $\sigma_{\ell}^2$ not to be too large compared to the mean.
%

Using the WKB ansatz  $P\simeq\exp\left(-K_0S\right)$ in Eq. \eqref{eq:ME-full}, we arrive at a Hamilton-Jacobi equation with   Hamiltonian:
\begin{equation}
H=\rho\left(e^{p_{\rho}}-1\right)+\frac{\rho^{2}}{1+\xi}\left(e^{-p_{\rho}}-1\right)+\frac{1}{2}\left(K_0{\cal B}_\ell-\nu\xi\right)\left(e^{p_{\xi}}-1\right)+\frac{1}{2}\left(K_0{\cal B}_\ell+\nu\xi\right)\left(e^{-p_{\xi}}-1\right),
\end{equation}
where $\rho=N/K_0$, $S$ is the action, and $p_{\rho}=\partial S/\partial \xi$ and $p_{\xi}=\partial S/\partial \xi$ are the momenta. Assuming $p_{\xi}\ll1$ and taking  ${\cal B}_{\rm B}$ from Eq. (4) one gets the following Hamiltonian:
\begin{equation}
H=\rho\left(e^{p_{\rho}}-1\right)+\frac{\rho^{2}}{1+\xi}\left(e^{-p_{\rho}}-1\right)-\nu\xi p_{\xi}+\frac{K_{0}\nu}{2\beta}\left(1-\xi^2\right)p_{\xi}^{2}.
\end{equation}
In the fast-varying  limit we can assume that the noise instantaneously equilibrates to a $\rho$- and $p_\rho$-dependent value, such that $\dot{\xi}=\partial H / \partial p_{\xi}\simeq0$, $\dot{p_{\xi}}=-\partial H/ \partial \xi\simeq0$.
Performing this adiabatic elimination, these equations yield the following effective Hamiltonian:
\begin{equation}
H=\rho\left(e^{p_{\rho}}-1\right)+\rho^{2}\left(e^{-p_{\rho}}-1\right)+\frac{K_{0}}{2\nu\beta}\rho^{4}\left(e^{-p_{\rho}}-1\right)^{2}.
\end{equation}
By solving for $H=0$, the stationary action is found to be $S=\rho\ln\rho-\rho+K_{0}/(12\nu\beta)\left(3\rho^{2}-2\rho^{3}\right)$,
which allows us to obtain the PDF of the total population size: $P\left(\rho\right)\simeq P_{\rm WKB}\left(\rho\right)\simeq P_{0}\exp\left\{ -K_{0}S\right\}$, where $P_0$ is a normalization factor.
Finally, with Eq. (5) of the main text,
the fixation probability in the realm of this WKB approximation is given by
\begin{equation}
\label{eq:fixWKB}
 \phi\left(x_{0}\right) \simeq\frac{\int_{0}^{\infty}P_{\rm WKB}\left(\rho,\nu/s\right)\exp\left[K_{0}\rho\left(1-x_{0}\right)\ln\left(1-s\right)\right]d\rho}{\int_{0}^{\infty}P_{\rm WKB}\left(\rho,\nu/s\right)d\rho}.
\end{equation}
The integrals can be performed using a saddle-point approximation, yielding
%
\begin{equation}
\label{eq:fix2_WKB_SM}
\phi_{\ell} \simeq\exp\left[ - K_0\eta\left(1-\frac{ K_0 s \eta \sigma^{2}}{2\nu}\right)\right],
\end{equation}
where we have approximated $e^{-\eta}\equiv\left(1-s \right)^{1-x_0}\approx 1-\eta$, and $\beta\simeq 1/(2\sigma^2)$, valid for $\sigma\ll 1$.
Based on  similar calculations performed for the \emph{asymmetric} B-EN and Gamma-distributed EN, we have confirmed  that, in the regime $\nu/s\gg 1$, the fixation probability is given by Eq.~(\ref{eq:fix2_WKB_SM}) as long as $\sigma=\sigma_{\ell}\ll 1$. This result holds for any $\ell$-EN, provided that it can be approximated in the close vicinity of its mean  by a Gaussian distribution with standard deviation $\sigma_{\ell}$ \footnote[2]{for U-EN we got a similar expression as Eq. \eqref{eq:fix2_WKB_SM}, but with a subleading pre-factor of $3/4$ instead $1/2$ in the exponent. This small difference is due to the fact that U-EN can not be approximated by a Gaussian distribution, even for very small standard deviation.}.

Formula \eqref{eq:fix2_WKB_SM}
thus provides an accurate approximation of the $S$-fixation probability $\phi_{\ell}$, accounting for both EN and DN, in the regime of high varying  rate ($\nu/s\gg 1$), and for small variance ($\sigma_{\ell}\ll 1$).
In particular, Eq.~\eqref{eq:fix2_WKB_SM} demonstrates that the  exponent $\alpha_{\ell}$ such that $\ln{(\phi_{\ell}/\phi_{\ell}^{\infty})}\sim (s/\nu)^{\alpha_{\ell}}$ when $\nu/s\gg 1$, see Eq.~(9) and Fig.~3, is $\alpha_{\ell}\approx 1$
for all continuous $\ell$-EN considered here when $\sigma_{\ell}\ll 1$. Moreover, assuming again $\phi_{\ell}\approx  {\rm exp}(-s \langle N\rangle_\ell)$ when
 $\nu/s\gg 1$, \eqref{eq:fix2_WKB_SM} also implies that $\langle N\rangle_\ell \sim  K_{0}~e^{(s/\nu)^{\theta_{\ell}}}$ with $\theta_{\ell}\approx 1$ when
$\sigma_{\ell}\ll 1$ and $\nu/s \gg 1$, as reported in Fig.~\ref{fig:moments}(b).

\subsection{A3. SIMULATION METHODS}
%
\label{A3}

Under a continuously varying environment, the evolution is characterized by the coupling of the continuous SDE, Eq.~(3), with the discrete birth-death process, defined by Eqs. (1)-(2), governing the population dynamics in a static environment, see Sec.~A1. In addition to   the already discussed analytical intricacy, this  also poses a number of  challenges on how to perform reliable computer simulations of the system's dynamics.

The case of discrete EN results in an augmented multivariate birth-death process (see Eq.~\eqref{eq:ME2} and [32,33,54,59-61]) 
that can be simulated {\it exactly} by means of  Gillespie-like algorithms~[75, 76]. 
However, the situation is very different with continuous EN, since, to the best of our knowledge,
   no ``exact simulation methods'' of the dynamics are known. It is therefore necessary to resort to some approximation simulation scheme
of the  system's dynamics.
  Various approximate methods can be found in the literature,
  like tau-leap method [78] 
  and various extensions, see e.g. [79]
  , or the simulation of the SDEs associated to the diffusion approximation of  \eqref{eq:ME-full}.

 Here, we have focused on two approaches that consistently mirror reliably the system's dynamics. An approach consists of simulating directly the {\it fictitious birth-death process}  defined by Eq. \eqref{eq:ME-full}, and devised as the  discrete counterpart of the  coupling of
 SDE~(3) to (1) via (2) of the main text, see Eq. \eqref{eq:ME-full} and [77].
 This allows us to simulate the system's dynamics as that of an augmented  birth-death process using  Gillespie-like algorithms. This approach works well when the environment does not change too rapidly, but becomes inefficient in the limit $\nu\gg 1$.
 This difficulty is overcome by our second simulation method, which consists of regarding Eq. (3) of the main text as modelling a ``fictitious chemical reaction'' occurring at rate $\nu$: after an average time $1/\nu$
 the environmental variable $\xi_{\ell}$ is  updated
 by drawing its new value from the stationary distribution $p^*(\xi_{\ell})$. With this method, the population dynamics is thus simulated with a Gillespie-like algorithm~[76]
 , for the birth-death process  defined by (1) augmented by the ``fictitious chemical reaction'' to which it is coupled via (2).
 We have extensively tested and compared our two simulation methods,  finding consistent  accuracy between them
 and excellent agreement with analytical results in the limits $\nu \to 0,\infty$. In our comparisons, we  found that the  fictitious chemical reaction method is significantly more efficient than the other approach under fast-varying environments (speeding up simulations by a factor ${\cal O}(K_0^2)$ when $\nu \gg 1$), and we have used it in most of our simulations.

To further improve the efficiency of the simulations, in some figures $\phi_{\ell}$ is calculated by finding the PSD $P_{\ell}(N,\nu)$ from the histograms obtained by binning simulations data in the formula (5), rather than by  sampling a large number of long simulation realizations (fixation has to be reached). This is particularly useful in Fig.~2(e), where many parameter sets are used, and in Fig.~3, where the fixation probabilities should be accurate enough to achieve similarly accurate results in the rate of convergence of $\phi_\ell$ to the asymptotic fixation probability. We have verified that the difference between this method and simulating $\phi_{\ell}$ directly is negligible, see Fig.~\ref{fig:newFigS3}.

To find the asymptotic fixation probability $\phi_\ell^\infty$ in Figs. 3 of the main text and~\ref{fig:moments}(b) using Eq. (5), the asymptotic PSD $P_{\ell}(N,\nu\rightarrow\infty)$ was needed. In this regime the EN self-averages, so the PSD can obtained by binning simulations data with constant carrying capacity $K=\mathcal{K}_\ell$, i.e. with {\it effective} EN. This method enabled us to find the asymptotic fixation probability accurately without having to simulate the original EN, that switches extremely rapidly at $\nu\to\infty$.

\label{sec:SM4}
	\begin{figure}[t]
		\centering
		\includegraphics[width=0.45\linewidth]{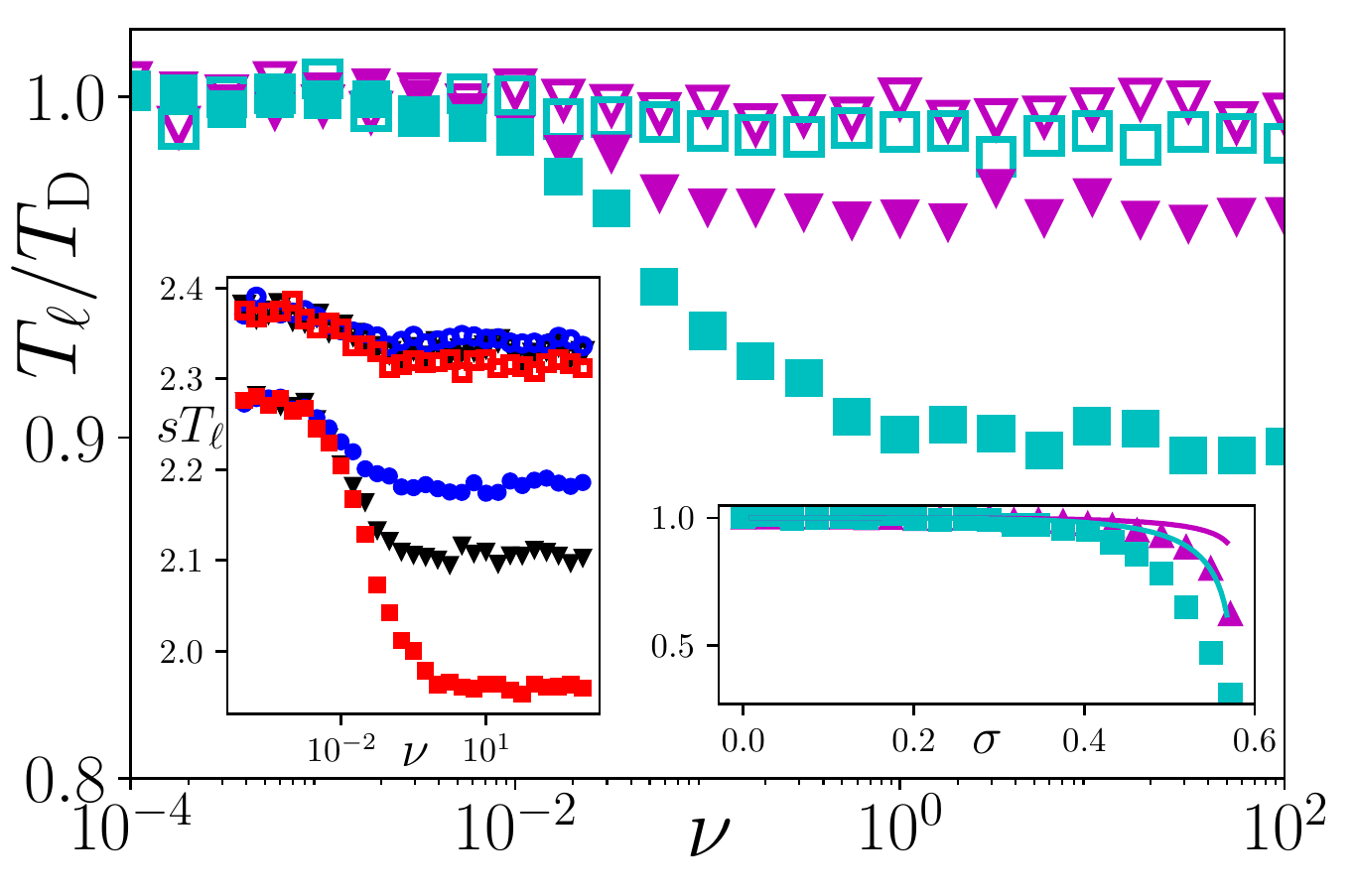}
		\vspace{-5mm}\caption{Main: Unconditional mean fixation time ratios: $T_{\rm U}/T_{\rm D}$ (magenta) and $T_{\rm B} / T_{\rm D}$ (cyan) vs. $\nu$, open (closed) symbols show results for $\sigma_{\ell}\equiv\sigma = 0.298$ $(0.44)$. Other parameters are $\left(s, K_0,x_0\right)=\left(0.02,250,0.5\right)$. Mean fixation times are of the same order with $T_{\rm D} > T_{\rm U} > T_{\rm B}$. Left inset: $sT_{\ell}$ for $\ell = ({\rm D}, {\rm U}, {\rm B})$ shown by blue circles, black triangles and red squares respectively. $sT_{\ell} = {\cal O}(1)$. Right inset: $T_{\rm U}/T_{\rm D}$ (magenta) and $T_{\rm B} / T_{\rm D}$ (cyan) vs. $\sigma$ in the large-$\nu$ regime (here, $\nu=1000$). Symbols are from simulations while lines show $\ln({\cal K}_{\ell}) / \ln({\cal K}_{\rm D})$.
		}
		\label{fig:mft}
	\end{figure}

\subsection{A4. MEAN FIXATION TIME}
\label{A4}
Similarly to $\phi_{\ell}$, the unconditional mean fixation time, $T_{\ell}$ can be obtained by integrating its counterpart for a constant population size $N$, $T(N,s,x_0)$, over the PSD $P_{\ell}(N,\nu/s)$~[32,33]:
\begin{eqnarray}
 \label{eq:formulaMFT}
 T_{\ell}\simeq
 \int_{0}^{\infty}~P_{\ell}(N,\nu/s)~T(N,s,x_0)~dN,
\end{eqnarray}
where $T(N,s,x_0)\sim (\ln{N})/s$ for $s\ll 1$ and $Ns \gg 1$ [30]. 
For all forms of EN, we thus find $T_{\ell}={\cal O}(1/s)$ with prefactors can change significantly, especially in the regime of high varying  rate $\nu$ and when $\sigma_{\ell}$ approaches $\sigma_{\rm max}$. The main panel of Fig.~\ref{fig:mft} shows that $T_{\ell} / T_{\rm D}$ is of order ${\cal O}(1)$, with $T_{\rm D}>T_{\rm U}>T_{\rm B}$. The left inset
of Fig. \ref{fig:mft}  confirms that $sT_{\ell}={\cal O}(1)$, while the right inset shows
that $T_{\ell} / T_{\rm D}\approx \ln{{\cal K}_{\ell}} / \ln{{\cal K}_{\rm D}}$ when $\nu/s\ll 1$
and $\sigma\ll  \sigma_{\rm max}$. These results hence suggest that $T_{\ell}$ scales as $1/s$ and that
$T_{\ell }\sim (\ln{{\cal K}_{\ell}})/s$ in the regime of high varying  rate when $\sigma \ll  \sigma_{\rm max}$.
It is  therefore clear that the effect of environmental variability is much more striking on the fixation probability than on mean fixation time, compare Figs. \ref{fig:newFigS3} and \ref{fig:mft}.	
	
\begin{figure}[t!]	\centering
	\includegraphics[width=0.9\linewidth]{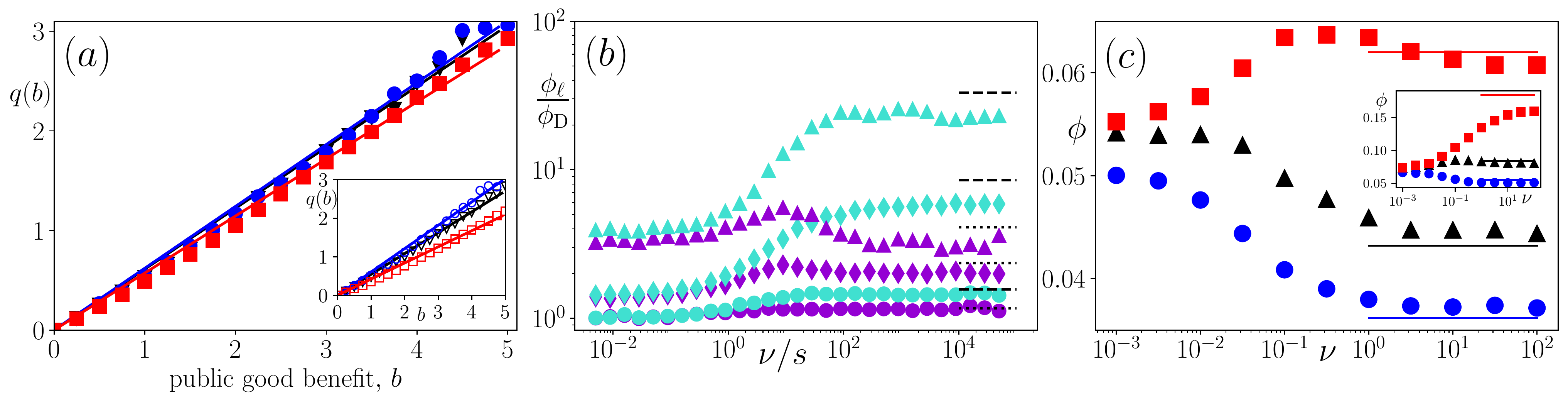}
	\vspace{-3mm}\caption{(a) $q_{\ell}(b)$ vs. $b$ for $\ell \in \mathcal{L}$ with $(s, K_0,x_0,\nu) = (0.02, 250, 0.5,1000)$. $\ell = {\rm D, U, B}$ represented by blue circles, black triangles and red squares respectively. Symbols are from simulations and lines show a best fit through $(0,0)$. Main (inset): $\sigma_{\ell} = 0.44$ $(0.519)$. $q_{\rm B} < q_{\lbrace \rm D, U \rbrace}$ for all $\sigma_{\ell}$, and $q_{\rm D} > q_{\rm U}$ for large $\sigma_{\ell}$ while $q_{\rm D} \simeq q_{\rm U}$ when $\sigma_{\ell}$ is small.  (b) $\phi_{\ell\in {\cal L}'}/\phi_{\rm D}$ (symbols)
	vs. $\nu/s$ with parameters $(s, K_0,x_0) = (0.02, 250, 0.5)$. Results for {\rm B}-EN/{\rm U}-EN are in cyan/purple; circles, diamonds, triangles respectively correspond to $(b,\sigma_{\ell})$ = $(0.5,0.44)$, $(2,0.519)$, $(4,0.519)$. Dashed lines show the prediction $\phi_{\rm B}/\phi_{{\rm D}}\approx \phi_{q_{\rm B}}/\phi_{q_{\rm D}}\approx  e^{\eta(\mathcal{K}_{\rm D}-\mathcal{K}_{\rm B})+\eta(q_{\rm D}\mathcal{K}_{\rm D}-q_{\rm B} \mathcal{K}_{\rm B})}>1$, while the dotted lines show the same for uniform noise.
	As predicted in panel (a), $\phi_{\rm B}/\phi_{\rm D} \geq \phi_{\rm U}/\phi_{\rm D}$, with the largest differences when $\sigma_{\ell}$ and $b$ are large.
 (c) $\phi_{\ell}$ vs. $\nu$ for $\ell={\rm B}$ (squares, red /gray), $\ell={\rm U}$ (triangles, black) and $\ell={\rm D}$ (circles, blue / dark gray) for $(b, s, K_0,x_0,\sigma_{\ell}) = (1, 0.02, 250, 0.5,0.44)$.
 Symbols are from simulations, red (gray)/black/blue (dark gray) horizontal lines show respectively
$\phi_{q_{\rm B}}, \phi_{q_{\rm U}}$ and $\phi_{q_{\rm D}}$ according to Eq.~\eqref{eq:phiq}. Inset shows the same with $\sigma_{\ell} = 0.519$.
}
%
	\label{fig:public_good}
\end{figure}

\subsection{A5. PUBLIC GOOD SCENARIO}
\label{sec:SM5}
 In the main text, we have focused on the the competition for resources
between slow and fast growers ($S$ and $F$ individuals), without any
explicit interactions between them.
This basic model is now generalized to account for public goods production and cooperative behaviour, which are issues of great biological relevance. Here, for simplicity we assume that slow growers ($S$ individuals) produce a public good (PG) shared with the entire population and benefiting equally PG producers as well as fast growers that can be seen as ``free riders'' since they exploit the PG without participating in its production. A proxy for the production of PG is therefore the fraction $x$ of PG-producers in the population,
and we simply assume that the birth rate of both strain is enhanced by a global term $g(x)=1+bx$, where
$b\geq 0$ and $b={\cal O}(1)$~[32,33,35,36,38,48]
, yielding the new birth rates $T_{S}^{+}= g(x)\frac{1-s}{\bar{f}}N_{S},\; T_{F}^{+}= \frac{g(x)}{\bar{f}} N_{F}$, while the death rates remain $T_{S}^{-}= \frac{N}{K_{\ell}}N_{S},\; T_{F}^{-}= \frac{N}{K_{\ell}}N_{F}$.

In the absence of any source of fluctuations, in the limit of an infinitely large population and assuming a constant carrying capacity $K_{\ell}=K_0\gg1$, the population’s mean-field dynamics obeys
\begin{equation}
\frac{dx }{dt} 
=
-\frac{s(1+bx)x(1-x)}{1-sx} \quad \text{and} \quad 	
		\frac{dN }{dt} 
		=N\left[1+bx-\frac{N}{K_0}\right].
\end{equation}
It is therefore clear that in this PG scenario ($b>0$) there is no timescale separation:
the dynamics of $x$ and $N$ are now
coupled, with $x$ still relaxing on a timescale $1/s$. Hence, when $0<s\ll 1$,
the fast variable $N$ is enslaved to the slowly-relaxing variable $x$.
This simple mean-field picture sheds light on the dynamics in this PG scenario when the population is finite population and subject to a fluctuating carrying capacity $K_{\ell}=K_{0}[1+\xi_{\ell}(t)]$: after a time of $t\sim 1/s$ there is fixation of either of the species;
$S$ and $F$ fixate with respective probabilities $\phi_{\ell}$ (then $x=1$) and $1-\phi_{\ell}$ (then $x=0$). After fixation, the population consists only of one species and is subject to either an effective carrying capacity $(1+b)K_{\ell}$ if $S$ fixated (in which case $x=1$), or $K_{\ell}$ if $F$ fixated (then $x=0$).
As a result of the coupling between $x$ and $N$, here the quasi-stationary  PSD $P_{\ell}(N,\nu,b)$ depends on which species fixates
according to
\begin{equation}
\label{eq:PSDb}
P_{\ell}(N,\nu,b)=\phi_{\ell}P_{\ell}(N,\nu)|_{K=(1+b)K_{\ell}}+(1-\phi_{\ell})P_{\ell}(N,\nu)|_{K=K_{\ell}},
\end{equation}
where $P_{\ell}(N,\nu)|_{(1+b)K_{\ell}}$
is the PSD obtained in the absence of PG production
conditioned to fixation of $S$ (hence weighted by $\phi_{\ell}$) subject to the effective
carrying capacity $(1+b)K_{\ell}$, whereas  $P_{\ell}(N,\nu)|_{K_{\ell}}$ is the  PSD when $b=0$ conditioned to fixation of $F$ (hence weighted by $1-\phi_{\ell}$)  subject to the
carrying capacity $K_{\ell}$~[32,33,59]
. This reflects the fact that when PG producers (strain $S$) fixates, which occurs with a probability  $\phi_{\ell}$, the population can grow bigger since
there are more resources available (PG is longer be exploited by free-riders, $K=(1+b)K_{\ell}>K_{\ell}$) than when  free riders $F$ fixate. The properties of the PSD shown in Fig.~\ref{fig:pdfs}(d,e,f) are obtained by
combining the PSDs of Fig.~\ref{fig:pdfs}(a,b,c)
according to \eqref{eq:PSDb}. As the result of the superposition \eqref{eq:PSDb} of the conditional PSDs, the PSD $P_{\ell}(N,\nu,b)$ under continuous EN is either bimodal in the large-$\nu$ regime, or exhibits a two-shoulder ($\ell={\rm B}$) or  two-step shape ($\ell={\rm U}$) in the slow/intermediate-$\nu$ regime. In contrast, $P_{{\rm D}}(N,\nu,b)$ can have up to four peaks in the slow-varying regime~[32,33].

The dynamics of this PG model prior to fixation is complicated by the absence of timescale separation and by the coupling of $N$ and $x$. However, analytical progress can be made in the regime of high varying rate
by extending the effective theory developed in Refs.~[32,33,59]
. Guided by the fact that the model's dynamics in the absence of EN is well described by a population of effective size, we introduce a parameter $q_{\ell}$, where $0\leq q_{\ell}\leq b$ and replace $g(x)$ by $1+q_{\ell}(b)$. The effective parameter $q_{\ell}$ is calculated by matching the fixation probability in the large-$\nu$ limit with $\phi(\mathcal{K}_{\ell}(1+q_{\ell}),s,x_0)$, i.e. by solving
\[
\phi_{\ell}=\frac{e^{-(\mathcal{K}_{\ell}(1+q_{\ell})x_0\ln(1-s)}-1}{e^{-\mathcal{K}_{\ell}(1+q_{\ell})\ln(1-s)}-1}\approx e^{\mathcal{K}_{\ell}(1+q_{\ell})(1-x_0)\ln{(1-s)}}
,\]
where the left-hand-side is obtained from simulation data obtained for large values of $\nu$ and $\mathcal{K}_{\ell}$ is given by Eq.~(8) in the main text.
Having determined $q_{\ell}$, see Fig.~\ref{fig:public_good}(a), we can effectively decouple
 $N$ and $x$ and calculate $\phi_{\ell}$ as in the case $b=0$, with the rescaled carrying capacity $K_0 \to K_0(1+ q_{\ell})$. Hence, in the regime of high varying  rate $\nu/s\gg 1$,  where this effective theory is expected to provide a good approximation for $\phi_{\ell}$, with $\eta$, we have
\begin{equation}
\label{eq:phiq}
\phi_{q_\ell}\approx e^{-\eta(1+q_{\ell})\mathcal{K}_{\ell}},
\end{equation}
 where the PG parameters $q_\ell$ are reported in Fig.~\ref{fig:public_good}(a)
 and found to grow essentially linearly with $b$, all the other parameters being kept fixed, with $q_{\rm B}<q_{\rm U},q_{\rm D}$ for different values of  $\sigma_{\ell}=\sigma$. This, together with
 ${\cal K}_{\rm B}\leq {\cal K}_{\rm U}\leq {\cal K}_{\rm D}$ (see Eq.(8)), means that
 in the regime of high varying  rate, the $S$ fixation probability is higher under ${\rm B}$-EN compared to ${\rm U}$ and ${\rm D}$ noise, with a difference that increases with $b$ and $\sigma_{\ell}$, as illustrated in  Fig.~\ref{fig:public_good}(b). Furthermore, while $\phi_{\rm D}$ and $\phi_{\rm U}$ are of the same order, we find $\phi_{\rm D}<\phi_{\rm U}$ when $\sigma$ is large enough, see Fig.~\ref{fig:public_good}(b).
  In fact, in the regime of high varying  rate ($\nu/s\gg 1$), we have $\phi_{\ell}/\phi_{{\rm D}}\approx \phi_{q_\ell}/\phi_{q_{\rm D}}\approx  e^{\eta(\mathcal{K}_{\rm D}-\mathcal{K}_{\ell})+\eta(q_{\rm D}\mathcal{K}_{\rm D}-q_{\ell}\mathcal{K}_{\ell})}>e^{\eta(\mathcal{K}_{\rm D}-\mathcal{K}_{\ell})}\geq 1$ for $\ell\in {\cal L}'$.
  Clearly,
  $S$ is typically more likely to fixate under continuous EN than under  {\rm D}-EN. The effect of continuous EN  on the
 $S$-fixation probability  increases with $b$
 and
 is stronger in this PG scenario than in the absence of PG production ($b=0$), see Fig.~\ref{fig:public_good}(b).
 In Fig.~\ref{fig:public_good}(c)
 we show that the large-$\nu$ effective theory prediction \eqref{eq:phiq}
 provides a good agreement with simulation results as soon as $\nu/s\gg 1$, that is $\phi_{\ell}\approx \phi_{q_\ell} $ for $\nu>1$ in the example of Fig.~\ref{fig:public_good}(c).

After $t\gtrsim 1/s$, fixation of one species is likely to have occurred and
$P_{\ell}(N)=\phi_{\ell}P_{\ell, S}(N)+(1-\phi_{\ell})P_{\ell, F}(N)$, where
$P_{\ell, S}(N)$ is the  PSD conditioned to fixation of $S$ (hence weighted by $\phi_{\ell}$) with effective
carrying capacity $(1+b)K_{\ell}$, whereas  $P_{\ell, F}(N)$ is the  PSD conditioned to fixation of $F$ (hence weighted by $1-\phi_{\ell}$) with effective
carrying capacity $K_{\ell}$.
As the result of the superposition of these conditional PSDs, we see in Fig.~\ref{fig:pdfs}(d,e,f) that for continuous EN,  the PSD $P_{\ell}(N)$ will generally be bimodal (or have two plateaus in the uniform noise case), whereas $P_{\rm D}(N)$
can have up to 4 peaks under slow switching~[32,33].


\begin{thebibliography}{56}
\bibitem{time_scales}
There is a third timescale, which is  exponential  in $K_0\gg 1$,  and corresponds to the extinction of the entire population. Notably, in Sec.~A5 of \cite{SM} we consider a public good scenario involving only the last two timescales.
\end{thebibliography}
\end{document}